\documentclass[11pt]{article}
\pdfoutput=1
\usepackage{jcapmod}
\usepackage{graphicx}
\usepackage{booktabs}
\usepackage[english]{babel}
\usepackage{amsmath,amssymb,amsbsy,amstext, amsthm, simplewick}
\usepackage{wrapfig}
\usepackage{amsfonts}
\usepackage{upgreek}
 \usepackage{exscale,relsize}
 \usepackage[makeroom]{cancel}
\usepackage{soul}
\usepackage{bbold}

\newcommand*\blob{\raisebox{-1.5pt}{\includegraphics[scale=0.6]{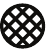}}}

\usepackage{colortbl}
\definecolor{lightgreen}{cmyk}{0.2, 0, 0.2, 0.2}

\def\be{\begin{equation}}
\def\ee{\end{equation}}
\def\bea{\begin{eqnarray}}
\def\eea{\end{eqnarray}}
\newcommand{\vs}{\nonumber\\}
\def\ba#1\ea{\begin{align}#1\end{align}}
\def\bg#1\eg{\begin{gather}#1\end{gather}}
\newcommand{\vev}[1]{\langle #1 \rangle}
\def\ph{\varphi}

\def\iMpch{\,h\,{\rm Mpc}^{-1}}

\newcommand{\refeq}[1]{Eq.~(\ref{eq:#1})}

\newcommand{\reffig}[1]{Fig.~\ref{fig:#1}}          
          
\newcommand{\refsec}[1]{Section~\ref{sec:#1}}  
\newcommand{\refssec}[1]{Sec.~\ref{sec:#1}}            
\newcommand{\refapp}[1]{App.~\ref{app:#1}}

\renewcommand{\v}[1]{{\boldsymbol{#1}}}

\newcommand{\vx}{\v{x}}

\newcommand{\vk}{\v{k}}
\newcommand{\vq}{\v{q}}
\newcommand{\vp}{\v{p}}
\newcommand{\<}{\langle}
\renewcommand{\>}{\rangle}

\renewcommand{\d}{\delta}

\newcommand{\eps}{\epsilon}

\def\P{\mathcal{P}}

\def\cH{\mathcal{H}}
\def\G{{\rm G}}

\newcommand{\comment}[1]{}

\makeatletter
\newlength{\apb@width}
\newcommand{\autoparbox}[2][c]{\settowidth{\apb@width}{#2}\parbox[#1]{\apb@width}{#2}}
\newcommand{\includegraphicsbox}[2][]{\autoparbox{\includegraphics[#1]{#2}}}
\makeatother

\def\xfl{\v{x}_{\rm fl}}

\def\der{\partial}

\def\vp{\v{p}}
\def\H{{\cal H}}

\def\psiE{\Psi}

\def\fnl{f_{\mathsmaller{\rm NL}}}

\def\Knl{K_{\mathsmaller{\rm NL}}}
\def\Gnl{G_{\mathsmaller{\rm NL}}}

\begin{document}

\begin{titlepage}

\setcounter{page}{1} \baselineskip=15.5pt \thispagestyle{empty}

\bigskip\

\vspace{1cm}
\begin{center}

{\fontsize{22}{24}\selectfont  \sffamily \bfseries  Galaxy Bias and Primordial Non-Gaussianity}

\end{center}

\vspace{0.2cm}

\begin{center}
{\fontsize{13}{30}\selectfont  Valentin Assassi$^{\bigstar,\spadesuit}$, Daniel Baumann$^{\bigstar,\clubsuit}$, and Fabian Schmidt$^\blacklozenge$}
\end{center}

\begin{center}

\vskip 8pt
\textsl{$^\bigstar$ DAMTP, University of Cambridge, CB3 0WA Cambridge, United Kingdom}
\vskip 7pt
\textsl{$^\spadesuit$ School of Natural Sciences, Institute for Advanced Study, NJ 08540, United States}
\vskip 7pt
\textsl{$^\clubsuit$ Institute of Physics, University of Amsterdam, 1090 GL Amsterdam, The Netherlands}
\vskip 7pt
\textsl{$^\blacklozenge$ Max-Planck-Institut f\"ur Astrophysik, 
85741~Garching, Germany}
\end{center}

\vspace{1.2cm}
\hrule \vspace{0.3cm}
\noindent {\sffamily \bfseries Abstract} \\[0.1cm]
We present a systematic study of galaxy biasing in the presence of primordial non-Gaussianity. 
For a large class of non-Gaussian initial conditions, we define a general bias expansion and prove that it is closed under renormalization, thereby showing that the basis of operators in the expansion is complete.   We then study the effects of primordial non-Gaussianity on the statistics of galaxies.
 We show that the equivalence principle enforces a relation between the scale-dependent bias in the galaxy power spectrum and that in the dipolar part of the bispectrum.  This provides a powerful consistency check to confirm the primordial origin of any observed scale-dependent bias.  
 Finally, we also discuss the imprints of anisotropic non-Gaussianity as motivated by recent studies of higher-spin fields during inflation.

\vskip 10pt
\hrule
\vskip 10pt

\vspace{0.6cm}
 \end{titlepage}

\tableofcontents

\newpage

\newpage
\section{Introduction}
\label{sec:intro}

Galaxy biasing is both a challenge and an opportunity.
On the one hand, it complicates the relation between the observed statistics of galaxies\footnote{Everything we say in this paper applies to arbitrary tracers of the dark matter density, even if we continue to refer to ``galaxies'' for simplicity and concreteness.} and the initial conditions. 
On the other hand, it may contain unique imprints of primordial non-Gaussianity (PNG)~\cite{Dalal:2007cu}. In this paper, we provide a systematic characterization of galaxy biasing for a large class of non-Gaussian initial conditions.

\vskip 4pt
At long distances, the galaxy density field can be written as a perturbative expansion
\be
\d_g(\vx,\tau) = \sum_O c_O(\tau)\, O(\vx,\tau)\,,
\label{eq:biasrel}
\ee
where the sum runs over a basis of operators $O$ constructed from the gravitational potential~$\Phi$ and its derivatives.  
For Gaussian initial conditions, the equivalence principle constrains the terms on the right-hand side of (\ref{eq:biasrel}) to be made from the tidal tensor $\partial_i \partial_j \Phi$.  A distinctive feature of
 primordial non-Gaussianity is that it can lead to apparently nonlocal correlations in the galaxy statistics. Moreover, the biasing depends on the soft limits of correlation functions which in the presence of primordial non-Gaussianity can have non-analytic scalings (i.e. $\propto k^\Delta$ in Fourier space, where $\Delta$ is not an even whole number).  
These effects cannot be mimicked by local dynamical processes and are therefore a unique signature of the initial conditions.

The bias expansion (\ref{eq:biasrel}) will contain so-called {\it composite operators}, which are products of fields evaluated at coincident points, such as $\delta^2(\vx,\tau)$.  In perturbation theory these operators introduce ultraviolet (UV) divergences in the galaxy correlation functions. Moreover, composite operators with higher spatial derivatives are not suppressed on large scales. Although these divergences can be regulated by introducing a momentum cutoff $\Lambda$, this trades the problem for a dependence of the galaxy statistics on the unphysical regulator $\Lambda$.  It is possible to reorganize the bias expansion in terms of a new basis of  {\it renormalized}  operators, $[O]$, which are manifestly cutoff independent~\cite{McDonald:2006mx,McDonald:2009dh,Assassi:2014fva,Schmidt:2012ys,Senatore:2014eva,Angulo:2015eqa}:
 \be
\d_g(\vx,\tau) = \sum_O b_O(\tau)\, [O](\vx,\tau)\, . 
\label{eq:biasrel2}
 \ee 
The basis of renormalized operators has a well-defined derivative expansion and the biasing model becomes an effective theory.

In this paper, we will explicitly construct the basis of operators in (\ref{eq:biasrel2}) for PNG whose bispectrum  in the squeezed limit has an arbitrary momentum scaling and a general angular dependence.
We will prove that our bias expansion is closed under renormalization, thereby showing that the basis of operators is complete. Completeness of the operator basis is a crucial aspect of the biasing model. Failing to account for all operators in the expansion could result in a misinterpretation of the primordial information contained in the clustering of galaxies.   On the other hand, a systematic characterization of the possible effects of late-time nonlinearities  allows us to identify observational features that are immune to the details of galaxy formation and hence most sensitive to the initial conditions.  
For example, we will show that the equivalence principle enforces a relation between the non-Gaussian contributions to the galaxy power spectrum and those of the dipolar part of the bispectrum, without any free parameters.  Combining these two detection channels for PNG provides a powerful consistency check for the primordial origin of the signal.  We also discuss the characteristic imprints of anisotropic non-Gaussianity as motivated by recent studies of higher-spin fields during inflation~\cite{Arkani-Hamed:2015bza} and of solid inflation~\cite{Endlich:2012pz}. 

Throughout, we will work in the standard quasi-Newtonian description
of large-scale structure~\cite{Bernardeau/etal:2002}.  One might wonder whether
there are relativistic corrections that, on large scales, become comparable to the scale-dependent
signatures of PNG that we will derive.  However, when interpreted in terms of proper time and distances, the quasi-Newtonian description remains valid on all scales \cite{CFC2}, and the only other scale-dependent signatures arise from photon propagation effects between source and observer, such as gravitational redshift. 


\vskip 10pt
The outline of the paper is as follows. In \refsec{RHB}, we introduce the systematics of galaxy biasing in the presence of PNG. We show that the bias expansion contains new operators which are sensitive to the squeezed limit of the  primordial bispectrum. We explicitly renormalize the composite operators $\delta^2$ and prove that our basis is closed under renormalization at the one-loop level.  Readers not concerned with the technical details can jump straight to \refssec{summary} for a summary of the results. 
In \refsec{bis}, we study the effects of these new operators on the statistics of galaxies. We derive a consistency relation between the galaxy power spectrum and the bispectrum, and determine the effects of anisotropic PNG on the galaxy bispectrum. Our conclusions are stated in \refsec{conclusions}.
Technical details are relegated to the appendices: 
In Appendix~\ref{app:systematics}, we derive a Lagrangian basis of bias operators equivalent to the Eulerian basis described in \refssec{basis}, and we extend the proof that the basis of operators is closed under renormalization to all orders.  
In Appendix~\ref{app:NG}, we study the effects of higher-order PNG.

\subsubsection*{Relation to Previous Work}

Our work builds on the vast literature on galaxy biasing which we shall briefly recall.  
A first systematic bias expansion, in terms of powers
of the density field, was introduced in \cite{Fry:1992vr} (this is frequently referred to as
``local biasing'').  The analog in Lagrangian space was studied 
for general initial conditions by \cite{Matarrese:1986et}.  The fact that local
Eulerian and local Lagrangian biasing are inequivalent was pointed out in~\cite{Catelan:2000vn}.
McDonald and Roy~\cite{McDonald:2009dh} addressed this at lowest order by including the tidal field (see also \cite{Chan:2012jj,Baldauf}), as well as higher-derivative terms, in the Eulerian bias expansion.  Finally, a complete basis of
operators was derived in \cite{Mirbabayi:2014zca,Senatore:2014eva}.  The need for renormalization of the bias parameters was first emphasized in \cite{McDonald:2006mx}, and further developed in \cite{Schmidt:2012ys,Assassi:2014fva, Senatore:2014eva}.   Scale-dependent bias was identified as a probe of PNG in \cite{Dalal:2007cu}, and further studied in \cite{Matarrese:2008nc,slosar/etal:2008,Schmidt:2010gw,Scoccimarro:2011pz,Desjacques:2011mq,Matsubara:2012nc, McDonald:2008sc}. 
A bivariate basis of operators was constructed in~\cite{Giannantonio:2009ak} (this is a subset of the basis we will derive in this paper).  Recently, this basis was used to derive the galaxy three-point function in the presence of local-type non-Gaussianity~\cite{Tellarini:2015faa}.   
The impact of anisotropic non-Gaussianity on the scale-dependent
bias was studied in~\cite{Raccanelli:2015oma}. Note that the derivation of~\cite{Raccanelli:2015oma} differs significantly from ours, since
it assumes a template for the bispectrum for all momentum configurations.  Moreover, it assumes that the dependence of
galaxies on the initial conditions is perfectly local in terms of the
initial density field smoothed on a fixed scale, which will not hold for realistic
galaxies.  In contrast, we will derive the bias induced by PNG in the squeezed limit, which is the
regime which is under perturbative  control (see also \cite{Angulo:2015eqa,Schmidt:2013nsa}).

\subsubsection*{Notation and Conventions}

We will use $\tau$ for conformal time and $\H$ for the conformal Hubble parameter.
Three-dimensional vectors will be denoted in boldface ($\vx$, $\vk$, etc.)~or with Latin subscripts ($x_i$, $k_i$, etc.).  The magnitude of vectors is defined as $k \equiv |\vk|$ and unit vectors are written as $\hat \vk \equiv \vk/k$. We sometimes write the sum of $n$ vectors as $\vk_{1\ldots n} \equiv \vk_1 + \ldots + \vk_n$. 
We will often use the following shorthand for three-dimensional momentum integrals
$$
\int_\vp\ (\ldots) \, \equiv\, \int \frac{{\rm d}^3 \vp}{(2\pi)^3}\, (\ldots)\ .
$$

We will find it convenient to work with the rescaled Newtonian potential $\Phi \equiv 2 \phi/(3\H^2 \Omega_m)$, so that the Poisson equation reduces to $\nabla^2 \Phi  = \delta$, where $\delta$ is the dark matter density contrast. 
A key object in the bias expansion is the tidal tensor $\Pi_{ij} \equiv \partial_i \partial_j \Phi$. Sometimes we will subtract the trace and write $s_{ij} \equiv \partial_i \partial_{j} \Phi - \frac{1}{3}\delta_{ij} \nabla^2\Phi $.  
We will use $\varphi$ for the primordial potential.
A transfer function $T(k,\tau)$ relates $\varphi(\vk)$ to the
 linearly-evolved potential and density contrast, 
 \begin{align}
 \Phi_{(1)}(\vk,\tau) &= T(k,\tau)\hskip 1pt \varphi(\vk)\, , \\
 \d_{(1)}(\vk,\tau) &= M(k,\tau)\hskip 1pt \varphi(\vk)\, , \label{eq:Mdef}
 \end{align}
 where $M(k,\tau)\equiv -k^2 T(k,\tau)$. 
 The linear matter power spectrum will be denoted by 
 \be
 P_{11}(k;\tau)\equiv\<\delta_{(1)}(\vk,\tau)\hskip 1pt\delta_{(1)}(-\vk,\tau)\>' \,=\, M^2(k,\tau) P_\varphi(k) \, ,
 \ee  
 where $P_\varphi(k) \equiv \<\varphi(\vk) \varphi(-\vk)\>'$.
 The prime on the correlation functions, $\vev{\cdots}'$, indicates that an overall momentum-conserving delta function is being dropped.
For notational compactness, we will sometimes absorb a factor of $(2\pi)^3$ into the definition of the delta function, i.e.~$\hat \delta_D \equiv (2\pi)^3 \delta_D$. 
 Non-Gaussianities in the primordial potential are  parametrized as 
\be
\varphi(\vk) = \varphi_\G(\vk) +\fnl\int_{\vp}\Knl(\vp,\vk-\vp)\big[\varphi_\G(\vp)\varphi_\G(\vk-\vp)-P_\G(p)\,\hat\delta_D(\vk)\big] + \cdots\ ,
\label{eq:NGexpX}
\ee
 where $\varphi_\G$ is a Gaussian random field and $P_\G(k)\equiv \<\varphi_\G(\vk)\varphi_\G(-\vk)\>'$. 
  At leading order in $\fnl$, this gives rise to the following primordial bispectrum 
   \begin{align}
B_\varphi(k_1,k_2,k_3) &\equiv\vev{\varphi(\vk_1)\varphi(\vk_2)\varphi(\vk_3)}'\nonumber\\[4pt]
 &=2\fnl \hskip 1pt \Knl(\vk_1,\vk_2)\hskip 1pt P_\varphi(k_1) P_\varphi(k_2)+ \text{2 perms}\, .
\end{align}
As we will see, the bias parameters are sensitive to the squeezed limit of the bispectrum. In this limit, and assuming a scale-invariant bispectrum, the kernel function in (\ref{eq:NGexpX}) can be written as
 \be
\Knl(\vk_\ell,\vk_s) \  \xrightarrow{\, k_\ell \ll k_s\, } \   \, \sum_{L,i} a_{L,i} \left(\frac{k_\ell}{k_s}\right)^{\Delta_i} \P_L(\hat \vk_\ell \cdot \hat \vk_s)\, ,
\label{eq:FNLSLX}
\ee
where $\P_{L}$ is the Legendre polynomial of even order $L$. 
 We call $\Delta_i$ and $L$ the scaling dimension(s) and the spin of the squeezed limit, respectively.
 
 \newpage
\section{Galaxy Bias and Non-Gaussianity}
\label{sec:RHB}

In this section, we will derive the leading terms of the biasing
expansion and describe the renormalization procedure for both Gaussian and non-Gaussian initial conditions. Readers who are less interested in the details of the systematic treatment of biasing can find a summary of our results in \refssec{summary}.

\subsection{Biasing as an Effective Theory}
\label{sec:EFThb}

\begin{figure}[t]
\centering
\includegraphics[width=0.7\textwidth]{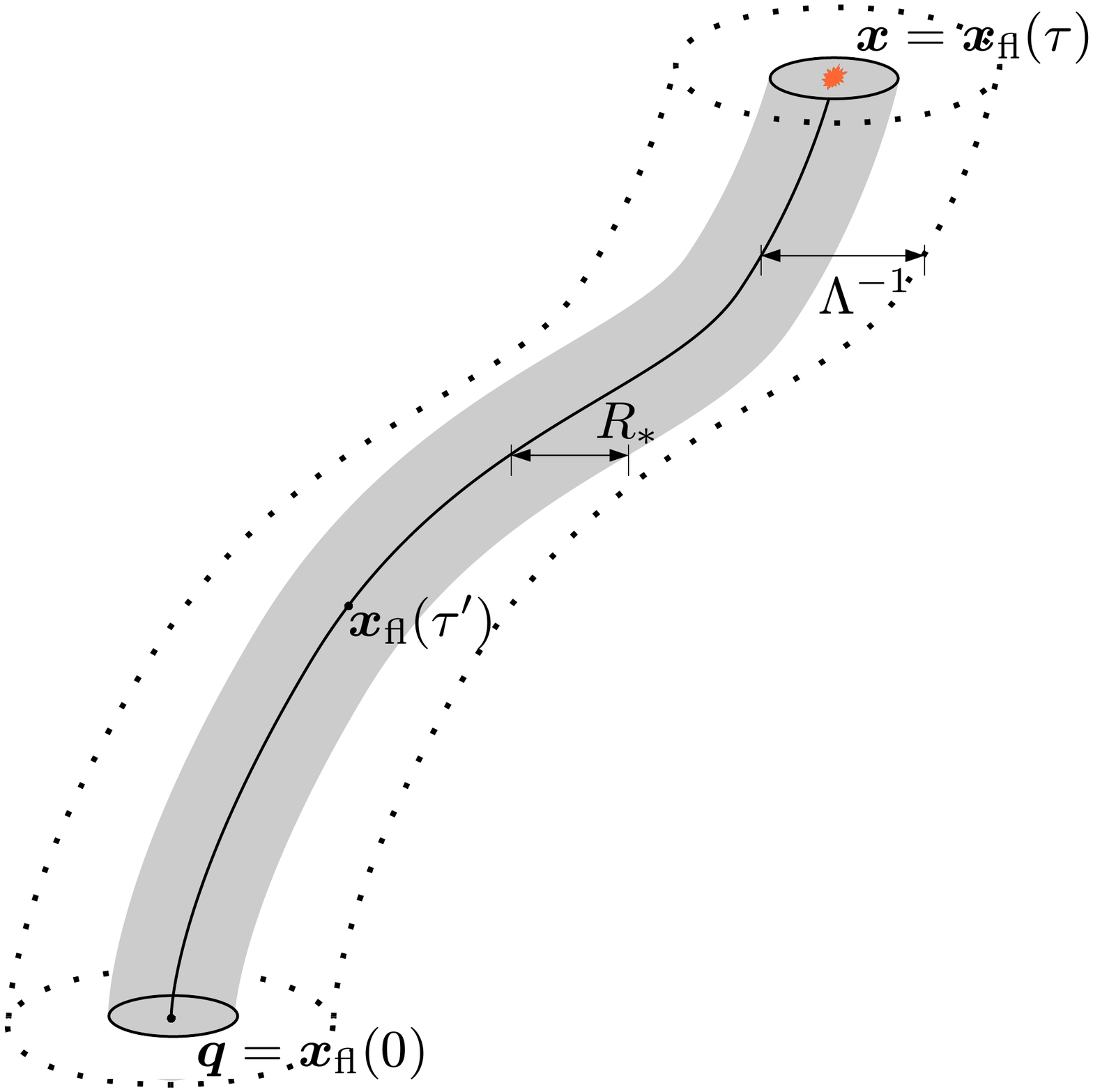}
\caption{Sketch of the spacetime region involved in the formation of galaxies. 
\label{fig:CFCsketch}}
\end{figure}

The number density of galaxies at Eulerian position $\vx$ and time $\tau$
is, in complete generality, a nonlinear and \emph{nonlocal} functional of the primordial potential
perturbations $\varphi(\v{y})$:  
\be
n_g(\vx,\tau) = {\cal F} \big[\varphi\big](\vx,\tau)\,.
\label{eq:nh0}
\ee 
Expanding this functional is not very helpful, since it would
lead to a plethora of free functions instead of a predictive bias
expansion.  To simplify the description we use the equivalence principle. This states that only second derivatives of the metric correspond to locally observable gravitational effects.  The bias expansion should therefore be organized in terms of the tidal tensor
\be
\Pi_{ij} \equiv 
 \partial_i \partial_j \Phi\, ,
\label{eq:Pidef}
\ee
where the spatial derivatives are with respect to the Eulerian coordinates.
We have used the rescaled potential in (\ref{eq:Pidef}), 
so that $\d^{ij} \Pi_{ij} = \d$ is the
matter density perturbation.  To apply the equivalence principle, we transform to the free-falling frame along the fluid flow, i.e.~we perform a time-dependent (but spatially constant) boost for each fluid trajectory. This locally removes any uniform or pure-gradient
potential perturbations.  In the end, $n_g(\vx,\tau)$ will depend on the
value of $\Pi_{ij}$ along the entire past trajectory (see \reffig{CFCsketch}), so that \refeq{nh0} becomes 
\be
n_g(\vx,\tau) = {\cal F} \big[\Pi_{ij}(\xfl'(\tau'))\big]\,,
\label{eq:nh1}
\ee 
where $\xfl'(\tau')$ is the position of the fluid element which at time $\tau > \tau'$ is located at $\vx'$.
The primes on the coordinates on the right-hand side of (\ref{eq:nh1}) indicate that the functional in ${\cal F}$ is still nonlocal in space and time.  However, since ${\cal F}$ is written in terms of the leading local gravitational observables,
we expect the scale of spatial nonlocality, $R_*$, to be comparable
to the size of the galaxy itself (e.g.~the Lagrangian radius for halos). This is much smaller than the scales over which we want to describe
correlations.  

We can use this fact to our advantage, by splitting the perturbations into long-wavelength parts ($\ell$)
and short-wavelength parts ($s$) relative to a smoothing scale $\Lambda^{-1} > R_*$.
The  exact scale of this split will become irrelevant once we have renormalized the operators.  
Above the coarse-graining scale, the dependence of ${\cal F}$ on the long-wavelength modes 
becomes local in space, and we obtain
\be
n_{g,\ell}(\vx,\tau) = {\cal F}_\ell \big[\Pi^\ell_{ij}(\xfl(\tau')); 
P_\d(\vk_s|\xfl(\tau'))\,,\,\cdots\big]\,,
\label{eq:nh1b}
\ee 
where $P_\d(\vk_s|\xfl(\tau'))$ is the local power spectrum of the small-scale part of $\d = {\rm Tr}[\Pi_{ij}]$ measured at a certain point along the
fluid trajectory. 
The ellipsis stands for higher-point statistics of~$\Pi_{ij}(\vk_s)$ and higher derivatives of the long-wavelength fields.  After renormalization, the higher-derivative contributions will be suppressed above the scale $R_*$.
On the other hand, since there is no hierarchy in the time scales of the evolution of the short- and long-wavelength fluctuations, the number density $n_g$ may still depend on the large-scale fields along the entire fluid trajectory $\xfl(\tau')$.  As we will explain in more detail in \refssec{basis}, this dependence on the history of the long-wavelength mode can  be captured by time-derivative operators (see also~\cite{Mirbabayi:2014zca}).   These time derivatives only begin to appear explicitly at third order in $\Pi_{ij}$. 

Note that small- and long-wavelength modes, by construction, do not have any overlap in
Fourier space, so $n_{g,\ell}$ depends on the former only
through their local statistics.   
Moreover, for Gaussian initial conditions, the local statistics of the small-scale perturbations depend on the long-wavelength perturbations $\Pi_{ij}^\ell$ only
through mode-coupling in the gravitational evolution.  
In the case of primordial non-Gaussianity, on the other hand,
short and long modes are coupled in the initial conditions. This is the effect we are mainly interested in.

It is sufficient to write the dependence of $n_{g,\ell}$ on the 
small-scale statistics in terms of the initial conditions, since in perturbation theory the gravitational evolution of the small-scale statistics from early times to the time $\tau$ is captured by $\Pi_{ij}^\ell$. Equation (\ref{eq:nh1b}) then becomes
\be
n_{g,\ell}(\vx,\tau) = {\cal F}_\ell \big[\Pi^\ell_{ij}(\xfl(\tau'));  
P_{\d}(\vk_s|\vq) \,,\,\cdots\big]\,,
\label{eq:nh2}
\ee 
where $\v{q} \equiv \xfl(\tau=0)$ and  $P_{\d}(\vk_s|\vq)$ 
denotes the power spectrum of small-scale \emph{initial}
density perturbations in the vicinity of $\vq$.     
It will be important that the initial short-scale statistics are defined with respect to the Lagrangian coordinate $\vq$.   

On large scales, where perturbation theory is valid, we may expand the functional in~(\ref{eq:nh2}) in powers of the long-wavelength fields and their derivatives. 
At second order in the fluctuations and to leading order in derivatives, the overdensity of galaxies can then be written as   
\begin{align}
\delta_{g,\ell}(\vx,\tau)&\equiv \frac{n_{g,\ell}(\vx,\tau)}{\bar n_g(\tau)} -1\nonumber\\[6pt]
&= f_0 + f_\Pi^{ij}\hskip 1pt\Pi_{ij}^{\ell}(\vx,\tau) + f_{\Pi^2}^{ijkl}\hskip 1pt\Pi_{ij}^{\ell}(\vx,\tau)\hskip 1pt\Pi^{\ell}_{kl}(\vx,\tau)+\cdots \, ,
\label{eq:dh1}
\end{align}
where $\bar n_g\equiv \vev{n_g}$ is the average number density of galaxies and the coefficients of this expansion,  $f_{O}[ P_{\d}(\vk_s|\vq) ,\cdots ]$, depend on the initial short-scale statistics.

\vskip 6pt
\noindent
Let us make a few comments:
\begin{itemize}
\item For Gaussian initial conditions, the coefficients $f_{O}$ in (\ref{eq:dh1}) are uncorrelated with the long-wavelength fields and are therefore simply cutoff-dependent parameters. More precisely, using statistical homogeneity and isotropy, these coefficients can be written as 
\begin{align}
f_0&=c_0\, ,\\[3pt]
f_{\Pi}^{ij} &= c_{\delta}\hskip 1pt\delta^{ij}\, ,
\label{fij}\\
f_{\Pi^2}^{ijkl} &= c_{\delta^2}\hskip 1pt\delta^{ij}\delta^{kl}+\frac{1}{2}c_{\Pi^2}\left(\delta^{ik}\delta^{jl}+\delta^{il}\delta^{jk}\right) ,
\label{fijkl} 
\end{align}
where the coefficients $c_{O}(\Lambda)$ are the bare bias parameters. Substituting this into~(\ref{eq:dh1}), we recover the usual bias expansion 
\be
\delta_{g,\ell}(\vx,\tau)=c_0 + c_\delta\hskip 1pt\delta_\ell(\vx,\tau)+c_{\delta^2}\hskip 1pt\delta^2_\ell(\vx,\tau)+c_{\Pi^2}\hskip 1pt(\Pi^{\ell}_{ij}(\vx,\tau))^2+\cdots\, .
\label{eq:dhG}
\ee
\item For non-Gaussian initial conditions, the initial short-scale statistics depend (in general non-locally) on the long-wavelength fields. This dependence is inherited by the coefficients~$f_{O}$ in (\ref{eq:dh1}). 
\item When the statistics of the short scales is isotropic, the tensor structures of the coefficients in~(\ref{eq:dh1}) are constrained to be (products of) Kronecker delta tensors; cf.~Eqs.~(\ref{fij}) and~(\ref{fijkl}). However, as we will see in \refssec{PNG}, in the presence of anisotropic PNG this is no longer the case and the tensor structure of these coefficients can be more complicated. 
\item  The expansion~(\ref{eq:dh1}) contains products of fields evaluated at coincident points, such as~$\delta^2_\ell$ and $(\Pi_{ij}^{\ell})^2$. These composite operators are the ones which yield divergences when computing galaxy correlation functions at the loop level and are precisely the terms we will need to renormalize (see \refssec{renorm}). 
\end{itemize}

\subsection{Non-Gaussian Initial Conditions}
\label{sec:PNG}

Next, we will derive the additional terms in the bias expansion that arise for PNG.  
If the initial conditions are statistically homogeneous and isotropic, we can write the primordial potential $\varphi$ as follows~\cite{Schmidt:2010gw}  
\be
\varphi(\vk) = \varphi_\G(\vk) +\fnl\int_{\vp}\Knl(\vp,\vk-\vp)\big[\varphi_\G(\vp)\varphi_\G(\vk-\vp)-P_\G(p)\,\hat \delta_D(\vk)\big] + \cdots\, ,
\label{eq:NGexp}
\ee
where $\varphi_\G$ is a Gaussian random variable and $P_\G(k)\equiv \<\varphi_\G(\vk)\varphi_\G(-\vk)\>'$. 
Throughout the main text, we will restrict to leading-order non-Gaussianities by truncating (\ref{eq:NGexp}) at second order.  This captures the effects of a primordial three-point function.
To account for primordial $N$-point functions, one should expand~(\ref{eq:NGexp}) up to order $(N-1)$ in~$\varphi_\G$. We will discuss the influence of higher-order PNG in Appendix~\ref{app:NG}.

\vskip 4pt
The primordial bispectrum associated with the quadratic term in (\ref{eq:NGexp})~is
\begin{align}
B_\varphi(k_1,k_2,k_3) &\equiv\vev{\varphi(\vk_1)\varphi(\vk_2)\varphi(\vk_3)}'\nonumber\\[4pt]
 &=2\fnl \hskip 1pt \Knl(\vk_1,\vk_2)\hskip 1pt P_\varphi(k_1) P_\varphi(k_2)+ \text{2 perms}\, .
 \label{eq:Bprimordial}
\end{align}
Note that the bispectrum does not uniquely specify the kernel~$\Knl$~\cite{Schmidt:2010gw,Scoccimarro:2011pz,Assassi:2015jqa}.  However, for non-singular kernels, the squeezed limit,
which is the relevant regime for biasing, is uniquely determined.  In this limit, we have
\be
\frac{B_\varphi(k_\ell,|\vk_\ell-\tfrac{1}{2}\vk_s|,|\vk_\ell+\tfrac{1}{2}\vk_s|)}{P_\varphi(k_\ell)P_\varphi(k_s)}\ \xrightarrow{\ k_\ell\ll k_s\ }\ 2\fnl\big[\Knl(\vk_\ell,\vk_s)+\Knl(\vk_\ell,-\vk_s)\big]\, . \label{eq:BB}
\ee
Statistical isotropy and homogeneity impose that the kernel function $\Knl(\vk_\ell,\vk_s)$ only depends on the magnitude of the two momenta, $k_\ell$ and $k_s$, and their relative angle $\hat\vk_\ell\cdot\hat\vk_s$. This angular dependence can conveniently be written as an expansion in Legendre polynomials. More precisely, we will assume that 
\be
\Knl(\vk_\ell,\vk_s) \  \xrightarrow{\, k_\ell \ll k_s\, } \   \, \sum_{L,i} a_{L,i} \left(\frac{k_\ell}{k_s}\right)^{\Delta_i} \P_L(\hat \vk_\ell \cdot \hat \vk_s)
\label{eq:FNLSL}\, ,
\ee
where $\P_{L}$ is the Legendre polynomial of even order $L$.\footnote{Since the squeezed limit in (\ref{eq:BB}) is invariant under $\vk_s\mapsto-\vk_s$, only Legendre polynomials of even order contribute to~(\ref{eq:FNLSL})~\cite{Assassi:2015jqa,Lewis:2011au, Shiraishi:2013vja}.} The ansatz (\ref{eq:FNLSL}) covers a wide range of inflationary models (e.g.~\cite{Chen:2009zp,Baumann:2011nk, Chen:2006nt, Alishahiha:2004eh, Green:2013rd, Arkani-Hamed:2015bza, Endlich:2012pz, Barnaby:2012tk, Shiraishi:2012sn,Shiraishi:2012rm}; see also~\cite{Angulo:2015eqa, Mirbabayi:2015hva}). 

\vskip 4pt
The squeezed limit of the bispectrum determines how the power spectrum of short-scale fluctuations is affected by long-wavelength fluctuations. To be more precise, consider the local short-scale power spectrum for a given realization of the large-scale fluctuations: 
\begin{align}
P_\ph(\vk_s|\vq)&\equiv \vev{\varphi_s(\vk_s)\varphi_s(-\vk_s)}'\big|_{\varphi^\ell_{\rm G}(\vq)}\nonumber\\
&= \left[1 + 4\fnl \int_{\vk_\ell} \Knl(\vk_\ell,\vk_s) \hskip 1pt\ph^\ell_{\rm G}(\vk_\ell) \hskip 1pt e^{i\vk_\ell\cdot \vq}\right] P_\ph(k_s)\, .
\label{eq:Ploc}
\end{align}
The
integral in (\ref{eq:Ploc}) only has support for $k_\ell<\Lambda$ and is sensitive to the squeezed limit of the kernel function. Substituting (\ref{eq:FNLSL}) into (\ref{eq:Ploc}), we find that the power spectrum receives contributions from each order (or ``spin'') of the Legendre expansion:  
\begin{itemize}
\item {\it Spin-0}

This is the well-known isotropic ($L=0$) contribution to the squeezed limit. For $\Delta=0$ and $\Delta=2$ this corresponds to local~\cite{Komatsu:2001rj} and equilateral~\cite{Chen:2006nt, Alishahiha:2004eh} non-Gaussianity, respectively. Intermediate values of $\Delta$ arise in inflationary models in which the inflaton interacts with light scalar fields~\cite{Chen:2009zp} or couples to operators of a conformal field theory~\cite{Green:2013rd}. 
Equation~(\ref{eq:Ploc}) then becomes
\be
P_\ph(\vk_s|\vq) = \Big[1 + 4\hskip 1pt a_0\fnl\hskip 1pt(\mu/k_s)^\Delta\psi(\vq)\Big] P_\ph(k_s)\, , 
\label{eq:Pph}
\ee
where we have defined the field 
\be
\psi(\vk) \equiv \left(\frac{k}{\mu}\right)^\Delta\varphi_\G^\ell(\vk)\, .
\label{eq:psidef}
\ee
The scale $\mu$ in (\ref{eq:Pph}) and (\ref{eq:psidef}) is an arbitrary reference scale. 
The non-dynamical field~$\psi$ parametrizes the dependence of the initial short-scale statistics on the long-wavelength field. This means that the coefficients of (\ref{eq:dh1}), which are functions of the initial short-scale statistics, depend on the field $\psi$. For example, at first order, the coefficients $f_0$ and $f_\Pi^{ij}$ in the expansion (\ref{eq:dh1}) are
\begin{align}
f_0&= c_0 +  c_{\psi}\hskip 1pt\psi(\vq) + \cdots\, ,\\
f_\Pi^{ij}&= \big[c_\delta +  c_{\psi\delta}\hskip 1pt\psi(\vq)\big]\hskip 1pt\delta^{ij} + \cdots\, ,
\end{align}
where the field $\psi$ is evaluated in Lagrangian space and the coefficients $c_i$ and $c_{i\psi}$ are the (cutoff-dependent) bare bias parameters. Defining  $\psiE(\vx,\tau)\equiv\psi(\vq(\vx,\tau))$, the bias expansion becomes 
\be
\delta_{g,\ell} =  c_0 + c_{\psi}\hskip 1pt\psiE +  c_\delta\hskip 1pt\delta_\ell+ c_{\psi\delta}\hskip 1pt\psiE\delta_\ell + c_{\delta^2}\hskip 1pt\delta^2_\ell +{c}_{\Pi^2}\hskip 1pt(\Pi_{ij}^\ell)^2+\cdots\, ,
\ee
where all the fields are implicitly evaluated at $(\vx,\tau)$. The field $\psiE(\vx,\tau)$ can be expanded in powers of the long-wavelength potential $\Phi_\ell$. At leading order, we have
\begin{align}
\psiE(\vx,\tau) 	&=\ \psi(\vx)\ +\boldsymbol{\nabla}\psi(\vx)\cdot\boldsymbol{\nabla}\Phi_\ell(\vx,\tau)+\cdots\, .
\label{eq:psiexp}
\end{align}
Note that the second term in this expansion involves a single derivative of the gravitational potential $\Phi$, which, by the equivalence principle, cannot appear on its own. In other words, this second term comes from the displacement of matter and is therefore constrained to only appear together with the first term $\psi(\vx)$.  

\vskip 4pt
Let us remark on the special case of equilateral PNG. Since the scaling in that case is $\Delta=2$, so that $\psiE \propto k^2 \varphi$,  the fields $\delta$ and $\psiE$ are indistinguishable on large scales. On small and intermediate scales, however,  $\d$ and $\psiE$ differ by a factor of the transfer function $T^{-1}(k)$. This may help to break the degeneracy between the two, although Gaussian higher-derivative operators will lead to similar scale dependences. We will discuss this further in \refssec{twopoint}.

\item {\it Spin-2} 

Considering the spin-2 contribution to (\ref{eq:FNLSL}), we find
\be
P_\varphi(\vk_s|\vq) = \Big[1+4\hskip 1pt a_2\fnl (\mu/k_s)^\Delta \,k_{s,i}k_{s,j} \hskip 1pt \psi^{ij}(\vq)\Big]P_{\ph}(k_s)\, ,
\ee
where
\be
\psi^{ij}(\vk)\equiv  {\cal P}^{ij}(\hat\vk)\hskip 1pt\psi(\vk) \, ,
\label{psiij}
\ee
with ${\cal P}^{ij}(\hat\vk)\equiv \frac{3}{2}(\hat k^i\hat k^j-\frac{1}{3}\delta^{ij})$.  We see that the small-scale power spectrum is now modulated by the tensor field $\psi^{ij}$.
At leading order, this leads to the following contribution to
the bias expansion 
\be
\delta_{g,\ell} \supset \psiE^{ij} \Pi_{ij}^\ell\, ,
\ee
where we have defined $\psiE^{ij}(\vx,\tau)\equiv \psi^{ij}(\vq(\vx,\tau))$.  As we will see in \refsec{bis}, such a term leaves a distinct imprint in the angular dependence of the galaxy bispectrum.  Note that for tensor observables, such as galaxy shapes, PNG with spin-2 contributes already at the two-point level \cite{Schmidt:2015xka}.

\item {\it Spin-4} 

Finally, the spin-4 contribution to the local short-scale power spectrum is
\be
P_\varphi(\vk_s|\vq) = \Big[1+4\hskip 1pt a_4\fnl (\mu/k_s)^\Delta \, k_{s,i}k_{s,j}k_{s,l} k_{s,m}\hskip 1pt \psi^{ijlm}(\vq)\Big]P_{\ph}(k_s)\, ,
\ee
where
\be
\psi^{ijlm}(\vk)\equiv {\cal P}^{ijlm}(\hat\vk)\psi(\vk)\, ,
\ee
and ${\cal P}^{ijlm}$ is a fully symmetric and traceless tensor (see~\cite{Assassi:2015jqa} for the precise expression).  However, at the order at which we are working, this term will not contribute.  Specifically, at lowest order in derivatives, the leading contribution to the bias expansion is a cubic term
\be
\delta_{g,\ell} \supset \psiE^{ijkl} \Pi_{ij}^\ell \Pi_{kl}^\ell\, .
\ee
At tree level, this only contributes to the trispectrum.
\end{itemize}

In the ansatz~(\ref{eq:FNLSL}), we have only considered the leading contribution to the primordial squeezed limit.  
The subleading corrections to the squeezed limit can be organized as a series in~$(k_\ell/k_s)^2$ \cite{Schmidt:2013nsa}.  
The next-to-leading term beyond the squeezed limit is then incorporated
in the bias expansion by the operator $\nabla^2\psi$, where derivatives are taken with respect to the Lagrangian coordinate. 
The bias coefficient of this term quantifies the response of the galaxy
number density to a change in the shape (rather than merely the amplitude) of the small-scale power spectrum.  We generically  expect these terms to be of the same order as higher-derivative operators in the bias expansion, which we will discuss in \refsec{conclusions}.  

\subsection{Systematics of the Bias Expansion}
\label{sec:basis}

We now describe how to systematically carry out the bias expansion
up to higher orders, starting from \refeq{nh2}.  
We will restrict ourselves to the lowest order in spatial derivatives,
which yields the leading operators on large scales.  Let us begin
by assuming Gaussian initial conditions.  
As discussed above, \refeq{nh2} still involves a functional dependence
on the long-wavelength modes along the past fluid trajectory.  
Consider a general operator~$O$ constructed out of the field\footnote{To avoid clutter in the expressions, we will drop the labels $\ell$ on the long-wavelength fields from now on.} $\Pi_{ij}^\ell \equiv \Pi_{ij}$.  At linear order, 
the dependence of $n_{g}(\vx,\tau)$ on $O$ can formally be written as
\ba
n_{g}(\vx,\tau) =\:& \int_{0}^\tau {\rm d}\tau'\:
f_O(\tau,\tau') \hskip 1pt O(\xfl(\tau'),\tau') 
\label{eq:nhe}\\
=\:& \left[ \int_{0}^\tau {\rm d}\tau'\:f_O(\tau,\tau') \right]
O(\vx,\tau)
+ \left[ \int_{0}^\tau {\rm d}\tau'\:f_O(\tau,\tau') (\tau'-\tau)\right]
\frac{\rm D}{{\rm D}\tau} O(\vx,\tau) + \cdots\,,
\nonumber
\ea
where ${\rm D}/{\rm D}\tau$ is a convective time derivative. 
In Eulerian coordinates, ${\rm D}/{\rm D}\tau$ is given  by 
\be
\frac{\rm D}{{\rm D}\tau} = \frac{\partial}{\partial\tau}
+ u^i \frac{\partial}{\partial x^i}\,,
\label{eq:DDtau}
\ee
where $u^i$ is the peculiar velocity.  
The expansion in (\ref{eq:nhe}) shows that we have to allow for convective time derivatives
such as ${\rm D}(\Pi_{ij})/{\rm D}\tau$, in the basis of operators.  
Including time derivatives of
arbitrary order then provides a complete basis of operators.  
Note, however, that the higher-order terms in the expansion (\ref{eq:nhe})
are not suppressed, since both galaxies and matter fields evolve over a Hubble time scale.  Fortunately, it is possible to
reorder the terms in (\ref{eq:nhe}) so that only a finite number
need to be kept at any given order in perturbation theory \cite{Mirbabayi:2014zca}.  

To do this, we do not work with the convective time derivatives
of operators directly, but instead take special linear combinations. These linear combination are chosen in such a way that
the contributions from lower-order operators cancel.  Let us denote operators that {\it start} at $n$-th order in perturbation theory with a superscript $[n]$, while $n$-th order {\it contributions} to
an operator are denoted with a superscript $(n)$. Consider the first-order contribution $\Pi^{(1)}_{ij}$
to $\Pi^{[1]}_{ij} \equiv \Pi_{ij}$. 
Taking the convective derivative of $\Pi^{(1)}_{ij}$ 
with respect to the logarithm of the growth factor $D(\tau)$, we have
\be
\frac{\rm D}{{\rm D}\ln D} \Pi^{(1)}_{ij} = (\cH f)^{-1} \frac{\rm D}{{\rm D}\tau} \Pi^{(1)}_{ij}
= \Pi^{(1)}_{ij}\, ,
\ee
where $f \equiv d\ln D/d\ln a$ is the logarithmic growth rate.
Hence, the operator 
\be
\Pi^{[2]}_{ij} \equiv \left(\frac{\rm D}{{\rm D}\ln D} - 1\right) \Pi^{[1]}_{ij}\,,
\ee
involves the first time derivative of $\Pi_{ij}$, but starts at second order
in perturbation theory.  This can be generalized to a recursive definition
at $n$-th order \cite{Mirbabayi:2014zca}, 
\be
\Pi^{[n]}_{ij} \equiv \frac{1}{(n-1)!} \left[(\cH f)^{-1}\frac{\rm D}{{\rm D}\tau} \Pi^{[n-1]}_{ij} - (n-1) \Pi^{[n-1]}_{ij}\right] .
\ee
Allowing for all time derivatives of operators constructed out of $\Pi_{ij}$ 
in the bias expansion is then equivalent to including the operators  
$\Pi^{[n]}_{ij}$ in the expansion.  That is, an expansion up to a given order
should contain all scalars that can be constructed out of 
$\Pi^{[n]}_{ij}$ at that order (see \refeq{listP} below).  
Note that, as emphasized in \cite{Mirbabayi:2014zca}, the higher-order terms
$\Pi_{ij}^{[n]}$ are in general \emph{nonlocal}
combinations of $\Pi_{ij}$, although they
only comprise a small subset of all possible nonlocal operators.    
Only these specific nonlocal operators should be included in the bias
expansion.  

Finally, there is one more restriction.  The quantity
${\rm Tr}[\Pi^{[n]}]$ corresponds to convective time derivatives of the
Eulerian density perturbation.  By the equations of motion, this is related
to  a linear
combination of lower-order operators, so it can be excluded from the basis of operators for~$n>1$.  

Up to third order, we then have the following list of bias operators for
Gaussian initial conditions~\cite{Mirbabayi:2014zca}: 
\bea
{\rm 1^{st}} \ && \ {\rm Tr}[\Pi^{[1]}] 
\label{eq:listP} \\[3pt] 
{\rm 2^{nd}} \ && \ {\rm Tr}[(\Pi^{[1]})^2]\,,\  ({\rm Tr}[\Pi^{[1]}])^2 \nonumber\\[3pt] 
{\rm 3^{rd}} \ && \ {\rm Tr}[(\Pi^{[1]})^3 ]\,,\ {\rm Tr}[(\Pi^{[1]})^2] \hskip 1pt  {\rm Tr}[\Pi^{[1]}]\,,\ ({\rm Tr}[\Pi^{[1]}])^3\,,\ {\rm Tr}[\Pi^{[1]} \Pi^{[2]}]\,, \nonumber
\eea
where all operators are evaluated at the same Eulerian position and time $(\vx,\tau)$.  
This basis offers the advantage of having a close connection to the standard
Eulerian bias expansions, i.e.~the terms in the first two lines correspond
exactly to those written in \refeq{dhG}.  In \refapp{basis}, we also 
provide an equivalent basis in Lagrangian space.  

In the non-Gaussian case, we have to extend the basis (\ref{eq:listP}) by the field $\psi(\vq)$, which is a nonlocal operator of the \emph{initial} density field; cf.~\refeq{psidef}.  
Using the Eulerian field $\psiE(\vx,\tau) \equiv \psi(\vq(\vx,\tau))$, we get
\bea
{\rm 1^{st}} \ && \  \psiE  
\label{eq:listNG} \\[3pt] 
{\rm 2^{nd}} \ && \ {\rm Tr}[\Pi^{[1]}]\hskip 1pt \psiE  \nonumber  \\[3pt]
{\rm 3^{rd}} \ && \ {\rm Tr}[(\Pi^{[1]})^2] \hskip 1pt \psiE\,,\  ({\rm Tr}[\Pi^{[1]}])^2 \hskip 1pt\psiE\,, \nonumber
\eea
and so on, where again all operators are evaluated at $(\vx,\tau)$.  The Lagrangian counterpart of this basis involves $\psi$ rather than $\psiE$ and is given in \refapp{basis}.
In \refssec{renorm} and App.~\ref{app:renorm}, we will show that the basis of operators defined in~(\ref{eq:listP}) and (\ref{eq:listNG}) is closed
under renormalization.  The generalization to higher-order PNG is given in Appendix~\ref{app:NG}.   

For anisotropic non-Gaussianity, the previous basis needs to be extended.  
Specifically, for the case $L=2$, the small-scale statistics are modulated by a trace-free 
tensor~$\psi_{ij}(\vq)$.  The leading contributions to the bias expansion then are
\bea
{\rm 1^{st}} \ && - 
\label{eq:listNGa} \\[3pt] 
{\rm 2^{nd}} \ && \ \Pi^{[1]}_{ij} \psiE^{ij}  \nonumber \\[3pt]
{\rm 3^{rd}} \ && \ ({\rm Tr}[\Pi^{[1]}]) \hskip 1pt \Pi^{[1]}_{ij} \psiE^{ij} \,, \nonumber
\eea
and so on, where as before $\psiE_{ij}(\vx,\tau)\equiv\psi_{ij}(\vq(\vx,\tau))$ and all operators are evaluated at $(\vx,\tau)$.

\subsection{Stochasticity and Multi-Source Inflation}
\label{sec:stoch}

The relation between biased galaxies and the underlying dark matter density fluctuations is in general
stochastic.  Physically, this stochasticity describes the random modulations
in the galaxy density due to short-scale modes whose statistics are
uncorrelated over large distances.  Such stochasticity can be described by introducing a set of random variables $\eps_i(\vx)$ which are uncorrelated with the matter variables and only have zero-lag correlations in configuration space.  They are thus completely described by their
moments $\< (\eps_i)^n (\eps_j)^m \cdots \>$, $n+m>1$, with $\<\eps_i\>=0$, since any non-zero expectation
value can be absorbed into the mean galaxy density.  Let us restrict to Gaussian initial conditions
for the moment.  We can demand that the moments of $\epsilon_i$ only depend on 
the statistics of the initial small-scale fluctuations $\ph(\vk_s)$, with
$|\vk_s| \gtrsim \Lambda$.  The influence of these small-scale initial
conditions on the late-time galaxy density will then depend on the long-wavelength observables 
through the gravitational evolution of the initial conditions.  Thus, we need to allow for stochastic terms in combination with each of the
operators in the basis discussed in \refssec{basis}.
Counting the stochastic fields as linear perturbations, we have to add
four stochastic fields $\eps_i$ up to cubic order, namely 
\bea
{\rm 1^{st}} \ && \ \eps_0 
\label{eq:stochbasis} \\[3pt]
{\rm 2^{nd}} \ && \ \eps_\delta \hskip 1pt{\rm Tr}[\Pi^{[1]}] \nonumber \\[3pt]
{\rm 3^{rd}} \ && \ \eps_{\Pi^2}\hskip 1pt {\rm Tr}[(\Pi^{[1]})^2]\,,\  \eps_{\d^2} \hskip 1pt ({\rm Tr}[\Pi^{[1]}])^2\,. \nonumber
\eea
 Let us note that, in principle, one could also have stochastic terms of the form $\epsilon_{ij}\Pi^{ij}$. However, in position space, correlation functions of $\epsilon_{ij}$ are proportional to (products of) Kronecker delta tensors and Dirac delta functions. For this reason, the effects of these terms on the statistics of galaxies are indistinguishable from those written in (\ref{eq:stochbasis}).  Hence, the basis (\ref{eq:stochbasis}) fully captures the effects of stochastic noise terms. 

Let us now consider the non-Gaussian case, and study under what conditions
PNG induces additional stochastic terms.   
By assumption, the stochastic variables $\eps_i$ only
depend on the statistics of the small-scale initial perturbations.  
As long as the coupling between long and short
modes is completely captured by the relation (\ref{eq:Ploc}), all effects
 are accounted for in our non-Gaussian basis (\ref{eq:listNG}).  
In this case, \refeq{stochbasis} only needs to be augmented by terms
of the same type multiplied by $\Psi$, 
\bea
{\rm 1^{st}} \ && -  \label{eq:stochbasisNG} \\[3pt]
{\rm 2^{nd}} \ && \ \eps_{\Psi}\Psi  \nonumber
 \\[3pt]
{\rm 3^{rd}} \ && \ \eps_{\Psi\delta}\Psi\hskip 1pt {\rm Tr}[\Pi^{[1]}] \,. \nonumber
\eea
As we show in App.~\ref{app:singlefield}, this holds whenever the initial conditions are
derived from a single statistical field,
corresponding to a single set of random phases. 
This is the case for the ansatz in (\ref{eq:NGexp}).  

Now, consider the correlation of the amplitude of small-scale initial perturbations over
large distances.  This can be quantified by defining the small-scale potential perturbations $\ph_s(\vx)$ through a high-pass filter $W_s$.  
Writing $\ph_s(\vk) \equiv W_s(k) \ph(\vk)$ in Fourier space, where $W_s(k)\to 0$ for $k \ll \Lambda$, we obtain the following
two-point function of $(\ph_s)^2(\vk)$:
\begin{align}
\< (\ph_s)^2(\vk)\,(\ph_s)^2(\vk')\>' \ =\: \left(\prod_{i=1}^4 \int_{\vk_i} \right)
 &\ \hat\d_D(\vk-\vk_{12}) \,\hat \d_D(\vk'-\vk_{34})  \nonumber  \\[-6pt]
& \ \times \<\ph_s(\vk_1)  \ph_s(\vk_2) \ph_s(\vk_3) \ph_s(\vk_4)\>\,.
\label{eq:phSphS}
\end{align}
Note that the high-pass filters ensure that the integral effectively runs only over $k_i\gtrsim \Lambda$.  
Large-scale perturbations, however, do contribute to this correlation in the collapsed
limit of the four-point function, e.g.~if $|\vk_{13}| \ll k_i$.  
If the non-Gaussian potential $\varphi$ is sourced by a single degree of freedom, then the collapsed
limit of the four-point function is completely described by the squeezed
limit of the bispectrum: both limits can be trivially derived from \refeq{Ploc}.  In that case, there is no additional source of stochasticity.  

On the other hand, if the initial conditions are sourced by more
than one field, then in general the collapsed limit of the four-point function
is larger than expected from the squeezed limit of the bispectrum~\cite{Smith:2011if,Baumann:2012bc}.  In that case, primordial
non-Gaussianity induces an additional source of stochasticity, i.e.~a
significant contribution to \refeq{phSphS}.
This stochastic contribution will be
cutoff-dependent  and has to be renormalized by a
stochastic counterterm, $\hat\psi$, with the following properties
\be
\< \hat\psi(\vk) \varphi_\G(\vk') \>' = 0 \quad\mbox{and}\quad
\< \hat\psi(\vk) \hat\psi(\vk') \>' = P_{\hat\psi\hat\psi}(k)\,.
\ee
The field $\hat \psi$ then has to be added to the bias expansion.
Note that, unlike the Gaussian stochastic fields~$\eps_i$, the field
$\hat\psi$ is characterized by a non-analytic power spectrum rather
than a white noise spectrum.  This reflects the completely different physical effects
encoded by the two types of fields: while the fields $\eps_i$ capture the dependence of the galaxy
density on the \emph{specific realization} of the small-scale modes,
the field $\hat\psi$ describes the modulation of small-scale
modes by long-wavelength modes which are \emph{uncorrelated} with 
$\varphi_\G$.  In general, $P_{\hat\psi\hat\psi}(k) \neq P_{\psi\psi}(k)$.  
Up to third order (but to leading order in $\fnl$), the following terms need to be added to the bias expansion
\bea
{\rm 1^{st}} \ && \ \hat\Psi 
\label{eq:stochNGterm} \\[3pt]
{\rm 2^{nd}} \ && \ \hat\Psi\,{\rm Tr}[\Pi^{[1]}]\,,\  \eps_{\hat\Psi} \hat\Psi
 \nonumber \\[3pt]
{\rm 3^{rd}} \ && \ \hat\Psi\, {\rm Tr}[(\Pi^{[1]})^2]\,,\  \hat\Psi\, ({\rm Tr}[\Pi^{[1]}])^2\,,\  
\eps_{\hat\Psi\delta}  \hat\Psi \hskip 1pt {\rm Tr}[\Pi^{[1]}]\,, \nonumber
\eea
where, in analogy with $\psiE$, we have defined $\hat\Psi(\vx,\tau) \equiv \hat\psi(\vq(\vx,\tau))$.  
The consequences of these contributions to the statistics of galaxies will be discussed in \refssec{twopoint}.

\subsection{Closure under Renormalization}
\label{sec:renorm}

At nonlinear order, the bias expansion contains composite operators, i.e.~products of fields evaluated at the same point.
These operators lead to divergences which need to be renormalized. In this section, we discuss the renormalization of composite operators in the presence of primordial non-Gaussianities. We show that every term in the basis of operators derived in the previous section is generated, but no more terms (see also App.~\ref{app:renorm}, where we extend the proof to all orders).  

\subsubsection*{Gaussian Initial Conditions}
\label{sec:GIC}

We will first recap the renormalization of the simplest composite operator, $\delta^2$, for Gaussian initial conditions (see also~\cite{McDonald:2009dh, Schmidt:2012ys, Assassi:2014fva}). 
Consider the correlations of $\delta^2$ with $m$ copies of the linearly-evolved density contrast $\delta_{(1)}$:
\be
C_{\delta^2,m}(\vk,\vk_i) \,\equiv\, \vev{\delta^2(\vk)\delta_{(1)}(\vk_1)\cdots\delta_{(1)}(\vk_m)}'\, . 
\label{eq:C}
\ee
This object will contain divergences which we wish to remove by subtracting appropriate counterterms from $\delta^2$. This procedure leads to the renormalized operator $[\delta^2]$, whose correlations with the linear density field, i.e.~$C_{[\delta^2], m}(\vk,\vk_i)$, are finite.  To uniquely fix the finite part of the correlator, we impose that the loop contributions to (\ref{eq:C}) vanish on large scales~\cite{Assassi:2014fva}
\be
\lim_{k \to 0} C_{[\delta^2],m}^{\rm loop}(\vk,\vk_i) = 0\, . 
\label{eq:RC}
\ee
This renormalization condition  is motivated by the fact that linear theory becomes a better approximation as one approaches large scales. The loop corrections are computed most easily using Feynman diagrams (see e.g.~\cite{Bernardeau:2001qr}).  The $n$-th order density contrast~$\delta_{(n)}$ will be represented by a square ($\mathsmaller \square$) with $n$ incoming lines attached to it. A black dot ($\bullet$) with two outgoing lines will represent the linear matter power spectrum $P_{11}$, while a black dot with three outgoing lines will refer to the linear (primordial) bispectrum $B_{111}$. For more details on the Feynman rules used in this paper, we refer the reader to~\cite{Assassi:2015jqa}. In the following, we construct the renormalized operator~$[\delta^2]$ up to $m=2$. This is sufficient to describe the one-loop galaxy bispectrum. 
\begin{itemize}
\item  For $m=0$, the expectation value of $\delta^2$ depends on the unphysical cutoff, $\vev{\delta^2}' \equiv \sigma^2(\Lambda)$. This dependence can be removed by simply subtracting this constant piece
\be
[\delta^2] \equiv \delta^2-\sigma^2(\Lambda)\ , \quad{\rm with}\quad \sigma^2(\Lambda) \equiv\int_0^\Lambda\frac{{\rm d} p}{2\pi^2}\, p^2P_{11}(p)\, .
\ee
This first renormalization step is always implicitly done in the literature as it ensures that $\vev{\delta_g} = 0$ at the loop level. 

\item For $m=1$, we have the following one-loop contribution
\be	
C_{[\delta^2],1}^{\rm loop}(\vk, \vk_1) \ =\quad \raisebox{-0.77cm}{\includegraphics[scale=0.7]{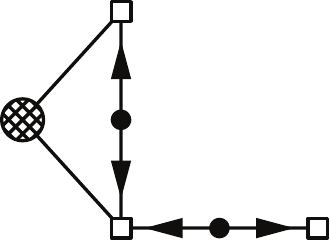}} \quad = \ \ \frac{68}{21}\hskip 1pt \sigma^2(\Lambda)\, P_{11}(k)\, ,
\ee
where the ``blob'' (\blob) 
in the Feynman diagram represents the operator $\delta^2$. 
We see that the loop diagram introduces a UV divergence proportional to $\sigma^2(\Lambda)$, which can be removed by defining the following renormalized operator
\be
[\delta^2]=\delta^2-\sigma^2(\Lambda)\left[1+ \frac{68}{21}\delta\right] .\label{eq:ct1}
\ee
\item Finally, considering $m=2$ diagrams, we have
\begin{align}
C_{[\delta^2],2}^{\rm loop}(\vk, \vk_1, \vk_2) \ \ &=\quad \raisebox{-0.77cm}{\includegraphics[scale=0.7]{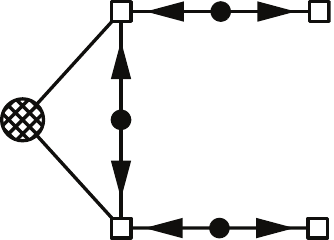}} \quad+\quad \raisebox{-1.27cm}{\includegraphics[scale=0.7]{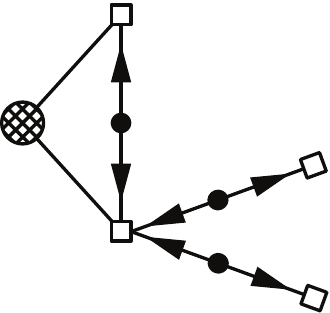}}\quad+\quad\raisebox{-0.48cm}{\includegraphics[scale=0.7]{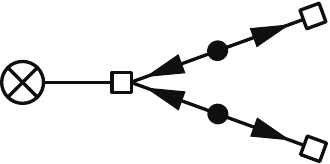}}\nonumber\\[10pt]
											&\xrightarrow{k_i\to0}\ 2\hskip 1pt \sigma^2(\Lambda)\left[\frac{2624}{735}+ \frac{254}{2205}(\hat\vk_1\cdot\hat\vk_2)^2\right]\,P_{11}(k_1)P_{11}(k_2)\, , 
\label{eq:m=2}
\end{align}
where $\otimes$ represents the $m =1$ one-loop counterterm.
The divergences in (\ref{eq:m=2}) can be absorbed by adding two more counterterms
\be
[\delta^2]=\delta^2-\sigma^2(\Lambda)\left[1+ \frac{68}{21}\delta+ \frac{2624}{735}\delta^2+ \frac{254}{2205}(\Pi_{ij})^2\right] . 
\label{eq:ct2}
\ee
\end{itemize}
This analysis can be extended straightforwardly to any composite operators and to higher loop order. 
In general, a renormalized operator $[O]$ is obtained by adding appropriate counterterms to the corresponding bare operator $O$:
\be
[{O}] \equiv {O} + \sum_{\tilde{O}} {Z}_{{O},\tilde {O}}(\Lambda) \tilde{O}\, ,
\ee
where the coefficients ${Z}_{{O},\tilde {O}}(\Lambda)$ are defined up to a finite (i.e.~cutoff-independent) contribution. 
The finite part is fixed by imposing the analog of the renormalization condition (\ref{eq:RC}). In terms of the new basis of renormalized operators, the bias expansion then becomes
\be
\d_g(\vx,\tau) = \sum_O b_O(\tau)\, [O](\vx,\tau)\, ,
\ee
where $b_{O}$ are the renormalized bias parameters. This expansion is manifestly cutoff independent, since both the renormalized operators and the renormalized bias parameters are independent of the cutoff. 

\subsubsection*{Non-Gaussian Initial Conditions}

In the presence of primordial non-Gaussianity, new diagrams appear as a result of a non-vanishing initial bispectrum and higher-point correlation functions. In this section, we will describe the renormalization of such diagrams. 
As before, we illustrate the renormalization procedure through the example of the simplest composite operator, $\delta^2$. 
We will show not only that the field $\psi$ defined in \refssec{PNG} is required to renormalize this composite operator, but also that it needs to be evaluated in Lagrangian space to ensure invariance under boosts. As before, the renormalization of composite operators is determined by looking at the divergences in the correlations with $m$ copies of the linearly-evolved dark matter density contrast $\delta_{(1)}$; see Eq.~(\ref{eq:C}). We consider the effects of the spin-0 and spin-2 contributions to the squeezed limit separately.

\vskip 6pt
\noindent{\it Spin-0.---}For an isotropic squeezed limit, we can guess the form of the renormalized operators before explicitly computing any non-Gaussian divergences. Indeed, in the presence of PNG, the short-scale variance $\sigma^2(\Lambda)$ is modulated by the field $\Psi$.  We therefore guess that the renormalized operators are simply obtained by replacing $\sigma^2(\Lambda)$ in the expression of the Gaussian renormalized operator (\ref{eq:ct2}) with
\be
\sigma^2(\Lambda) + a_0\fnl\sigma^2_\Delta(\Lambda)\psiE(\vx,\tau)\ , \qquad {\rm where} \qquad
 \sigma^2_\Delta(\Lambda)\equiv \int_0^\Lambda \frac{{\rm d} p}{2\pi^2}\,p^2 \left(\frac{\mu}{p}\right)^\Delta P_{11}(p)\ . 
 \label{eq:sigmamod}
\ee
Next, we will explicitly compute the non-Gaussian renormalized operator $[\delta^2]$ and show that this intuition is indeed correct.  

At one-loop order, the $m=0$ correlation function does not have a contribution from non-Gaussian initial conditions. We therefore start by looking at the $m=1$ correlation function.
\begin{itemize}
\item For $m=1$, the non-Gaussian contribution is
\begin{align}
\<[\delta^2](\vk)\delta_{(1)}(\vk')\>' 	&\ =\quad\raisebox{-0.77cm}{\includegraphics[scale=0.7]{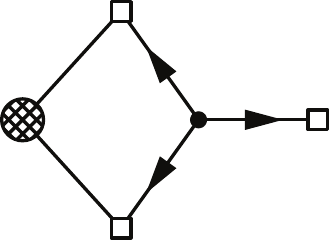}}\quad\nonumber\\&\ =\ \  \int_\vp B_{111}(p,|\vk-\vp|,k)\ \xrightarrow{k\to0}\ a_0\fnl\sigma^2_\Delta(\Lambda)\,P_{1\psi}(k)\, , 
\label{eq:m1div}
\end{align}
where $B_{111}$ is the linearly-evolved dark matter bispectrum. The cutoff-dependent function $\sigma^2_\Delta(\Lambda)$ was defined in~(\ref{eq:sigmamod})
\begin{align}
P_{1\psi}(k)&\equiv \<\delta_{(1)}(\vk)\Psi_{(1)}(\vk')\>'=\left(\frac{k}{\mu}\right)^\Delta \frac{P_{11}(k)}{M(k)}\, ,
\end{align}
where $\Psi_{(1)}\equiv\psi$ is the first-order contribution to the expansion~(\ref{eq:psiexp}) and $M(k)$ is the transfer function defined in (\ref{eq:Mdef}).
It is easy to see that the divergence in (\ref{eq:m1div}) is removed by a counterterm proportional to~$\Psi$:
\be
[\delta^2]^{\rm NG} = \delta^2 - a_0\fnl\sigma^2_\Delta(\Lambda) \Psi\, ,
\label{eq:ct1NG}
\ee
where the superscript ``NG'' reminds us that here we are only considering non-Gaussian counterterms. 

\item For $m=2$, the non-Gaussian diagrams are
\begin{align}
\<[\delta^2](\vk)\,\delta_{(1)}(\vk_1)\delta_{(1)}(\vk_2)\>' &= \quad \raisebox{-0.0cm}{\includegraphicsbox[scale=0.7]{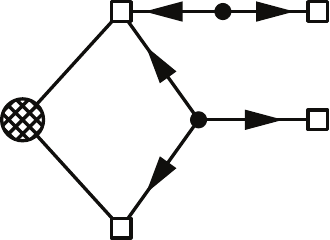}}\quad + \quad \includegraphicsbox[scale=0.77]{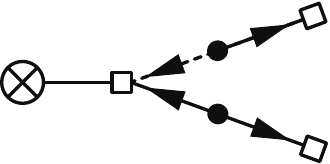}
\label{eq:div2graph}\\[10pt]
&\xrightarrow{k_i\to0} \ \  \frac{68}{21} a_0\fnl \sigma^2_\Delta(\Lambda)P_{11}(k_1) P_{1\psi}(k_2)\  + \{\vk_1\leftrightarrow\vk_2\}\, .
\label{eq:div2}
\end{align}
The semi-dashed line in the second diagram of~(\ref{eq:div2graph}) represents $P_{1\psi}$. This diagram 
arises from the second-order solution of the linear counterterm $\Psi$ in~(\ref{eq:ct1NG}), which comes from expanding the field $\Psi(\vx,\tau)=\psi(\vq)$ around $\vq= \vx$; cf.~Eq.~(\ref{eq:psiexp}). The divergence in (\ref{eq:div2}) is removed by a counterterm proportional to $\Psi\delta$:
\be
[\delta^2]^{\rm NG} = \delta^2 - a_0\fnl\sigma^2_\Delta(\Lambda)\left[1+ \frac{68}{21}\delta\right]\Psi\, .
\ee
\end{itemize}
Let us make a few comments:
\begin{itemize}
\item First, we note that while the individual diagrams in~(\ref{eq:div2graph}) yield non-boost-invariant divergences proportional to $\vk_i/k_i^2$, they cancel in the sum. This shows that it is crucial that the field $\psi$ is evaluated in Lagrangian space (otherwise the second diagram in (\ref{eq:div2graph})  would be missing). 
\item Including the leading Gaussian counterterm from (\ref{eq:ct2}), we find 
\be
[\delta^2] = \delta^2-\left[\sigma^2(\Lambda)+a_0\fnl\sigma^2_\Delta(\Lambda)\Psi\right]\left[1+\frac{68}{21}\delta\right] .
\ee
As anticipated, the renormalized non-Gaussian operators can be obtained by replacing the Gaussian variance of the short modes $\sigma^2(\Lambda)$ by the variance of the short modes modulated by the  long-wavelength fluctuations, i.e. $ \sigma^2(\Lambda) + a_0\fnl\sigma^2_\Delta(\Lambda)\Psi$.
\end{itemize} 
\noindent{\it Spin-2.---}Next, we consider the spin-2 contribution to the squeezed limit. As before, the non-Gaussian contribution to the $m=0$ divergence vanishes. Furthermore, looking at the $m=1$ correlation function, we find that the leading large-scale contribution (i.e.~$k\to0$) to the loop integral vanishes after angular integration: 
\begin{align}
\langle[\delta^2](\vk)\delta_{(1)}(\vk_1)\rangle' 	&= \int_\vp B_{111}(p,|\vk-\vp|,k)
\nonumber\\&
\xrightarrow{k\to0}\ a_2\fnl\sigma^2_\Delta(\Lambda)\int\frac{{\rm d}^2\hat\vp}{4\pi}\,\P_2(\hat\vk\cdot\hat\vp)P_{1\psi}(k)\,=\,0\, .
\end{align}
This was expected, since there cannot be a counterterm which is linear in $\Psi^{ij}$ (recall that $\Psi^{ij}$ is symmetric and traceless). However, the next-to-leading order contribution---i.e.~the one obtained by expanding the integrand to second order in $k/p$---comes with two additional powers of $k$ and is therefore renormalized by a higher-derivative term $\partial_i\partial_j\Psi^{ij}$.

\vskip 4pt
The $m=2$ correlation function has the following divergence 
\begin{align}
\langle[\delta^2](\vk)\,\delta_{(1)}(\vk_1)\delta_{(1)}(\vk_2)\rangle' &\ =\ \ \ \includegraphicsbox[scale=0.8]{delta22NG.pdf}\nonumber\\[10pt]
&\hspace{-1.5cm}\xrightarrow{k_i\to0}\ \frac{8}{105}a_2\fnl\sigma^2_\Delta(\Lambda)\left(3\hskip 1pt (\hat\vk_1 \cdot \hat \vk_2)^2 -1\right)P_{1\psi}(k_1) P_{11}(k_2)
+\{\vk_1\leftrightarrow\vk_2\}\, .
\label{eq:div2X}
\end{align}
We see that the term $\Psi^{ij} \Pi_{ij}$ is required to remove this divergence. More precisely, we have
\be
[\delta^2]^{\rm NG}= \delta^2 -\frac{16}{105}a_2\fnl \sigma^2_\Delta(\Lambda)\Psi^{ij}\Pi_{ij}\, .
\ee
We have therefore found that, up to second order, every term in the operator basis derived in \refssec{basis} is generated under renormalization, but no more terms. This suggests that our basis of operators is closed under renormalization. We prove this explicitly in Appendix~\ref{app:systematics}.

\subsection{Summary}
\label{sec:summary}

We carried out a systematic treatment of biasing and showed that the bias expansion can be written as a sum of a Gaussian and non-Gaussian contribution, 
$\delta_g=\delta_g^{\rm G}+\delta_g^{\rm NG}$, where $\delta_g^{\rm NG}$ contains all terms that scale as $\fnl$. Working at second order in fluctuations, we saw that the Gaussian contribution depends only on the tidal tensor $\Pi_{ij}\equiv\partial_i\partial_j\Phi$.  On the other hand, PNG gives rise to a modulation of the initial short-scale statistics by the long-wavelength perturbations. This is parametrized by a non-dynamical field $\Psi$ [cf.~\refeq{psidef}], which reduces to the primordial potential $\varphi$ for local PNG. If the squeezed limit of the bispectrum is anisotropic, this modulation is captured by tensor fields, such as $\Psi^{ij}$.  Furthermore, in cases where the initial potential perturbations are sourced by multiple fields, we have to allow for an additional field $\hat\Psi$ which captures the part of the long-short mode coupling that is uncorrelated with the long-wavelength potential itself.  

At second order in fluctuations and to leading order in derivatives, we find that the Gaussian and non-Gaussian contributions to the bias expansion are
\begin{align}
\delta_g^{\rm G}&= b_\delta\delta+b_{\delta^2}[\delta^2]+b_{s^2}[s_{ij}^2] + \eps_0 + [\eps_\d \d]
+ \cdots\, ,\\[4pt]
\delta_g^{\rm NG}&= \fnl\Big( b_\Psi\Psi+ b_{\Psi\delta}[\Psi\delta]+ b_{\Psi s }[\Psi^{ij} s_{ij}] + [\eps_{\Psi} \Psi] + \cdots \label{eq:deltaNG} \\
& \quad\quad\quad\   + b_{\hat\Psi} \hat\Psi + b_{\hat\Psi \delta} [\hat\Psi \d] + [\eps_{\hat\Psi} \hat\Psi] + \cdots \Big)\, ,
\nonumber
\end{align}
where $s_{ij}\equiv \Pi_{ij}-\frac{1}{3}\delta\hskip 1pt\delta_{ij}$ is the traceless part of the tidal tensor. Note that this expansion is written in terms of the renormalized operators (see \refssec{renorm}).   

\vskip 4pt
In contrast to the bare bias parameters, which depend on an arbitrary cutoff scale,
the renormalized bias parameters written in (\ref{eq:deltaNG}) have
 well-defined physical meanings.  For example, the density bias
parameters $b_{\d^n}$ correspond to the response of the galaxy abundance
to a change in the background density of the universe \cite{Cole:1989vx,Mo:1995cs,Baldauf:2011bh,Jeong:2011as,Schmidt:2012ys}
\be
b_{\delta^{n}} = \frac1{n!} \frac{\bar\rho^{\hskip 1pt n}}{\bar n_g} \frac{\partial^n \bar n_g}{\partial\bar\rho^{\hskip 1pt n}}\, .
\ee
Similarly, $b_{s^2}$ corresponds to the change in $\bar n_g$ due to an infinite-wavelength tidal field.  The non-Gaussian bias parameter $b_\Psi$, on the other hand, quantifies the response of $\bar n_g$ to a specific change in the 
primordial power spectrum amplitude and shape \cite{slosar/etal:2008,Schmidt:2010gw,Schmidt:2012ys},  
\be
b_\Psi =   \frac{1}{\bar n_g} \frac{\partial \bar n_g}{\partial \fnl} \, ,
\quad\mbox{where}\quad
P_{11}(k,\fnl) = \left[1 +4a_0\fnl\left(\frac{\mu}{k}\right)^\Delta\right] P_{11}(k)\,  . 
\label{eq:b_Psi}
\ee
Note that $b_\Psi$ is a function of the scaling dimension $\Delta$ (in addition to $a_0$) and that the dependence on the scale $\mu$ cancels with that in the definition of $\psi$.  
Correspondingly, $b_{\Psi \d}$ quantifies the response of $\bar n_g$ to a 
simultaneous change in the background density and the primordial power
spectrum.  The bias parameter $b_{\Psi s}$ describes the response of $\bar n_g$
to a combined long-wavelength tidal field and an anisotropic initial
power spectrum (see also the discussion in \cite{Schmidt:2015xka}).  
Analogous relations hold for $b_{\hat\Psi}$ and $b_{\hat\Psi\d}$.  
Note that $b_{\hat\Psi} \propto b_{\Psi}$
if the fields that source the curvature perturbations have the same scaling dimensions, i.e.~$\hat \Delta = \Delta$.  

\vskip 4pt
Let us point out that quadratic terms such as $\Psi^2$ and $(\Psi_{ij})^2$, which contribute at second order in~$\fnl$, have been dropped in (\ref{eq:deltaNG}). This is because cubic non-Gaussianity, which we have neglected starting
from (\ref{eq:NGexp}), will contribute terms of similar order and would
thus have to be included as well, see Appendix~\ref{app:NG}. 
This goes beyond the scope of this paper.  Note that these terms only
become relevant in the galaxy three-point function when all momentum modes
are very small, i.e.~of order $\cH$.  Galaxy surveys will have very low
signal-to-noise in this limit for the foreseeable future.  
Finally, terms of spin equal to four (or higher) contribute only at higher order in fluctuations, derivatives or non-Gaussianity.  We will therefore focus on the spin-0 and spin-2 contributions only.

\section{Galaxy Statistics}
\label{sec:bis}

We now study the effects of the non-Gaussian terms in the bias expansion (\ref{eq:deltaNG}) on the statistics of galaxies. It is well known that a primordial bispectrum with a non-vanishing squeezed limit yields a boost in the large-scale statistics of galaxies~\cite{Dalal:2007cu}. 
We will reproduce this effect, but also identify an additional, correlated signature in the angular structure of the bispectrum.  

\subsection{Power Spectra} 
\label{sec:twopoint}

We start with a brief review of the effects of PNG on the  two-point functions of fluctuations in the galaxy and (dark) matter densities. 
All fields will be evaluated at the same time $\tau$ (or redshift~$z$), so we drop the time arguments in the following.

\vskip 6pt
The leading contribution to the galaxy-matter cross correlation comes from the terms $\delta$ and~$\Psi$ in the bias expansion: 
\begin{align}
P_{gm}(k)\ \equiv\ \<\delta_g(\vk)\delta(\vk')\>' 
		&\ =\ b_\delta P_{11}(k)+\fnl b_\Psi P_{1\psi}(k)\ \nonumber\\[4pt]
&\ =\ \left(b_\delta+\Delta b(k)\right)P_{11}(k)\, ,
\end{align}
where $\Delta b(k)$ is the scale-dependent contribution to the linear bias induced by the field $\Psi$~\cite{Baumann:2012bc}
\be
\Delta b(k)\equiv \fnl b_\Psi \frac{(k/\mu)^\Delta}{M(k)}\, .
\label{eq:deltab}
\ee
Correspondingly, the leading contribution to the galaxy-galaxy auto correlation is
\begin{align}
P_{gg}(k) 	&\ \equiv\ \<\delta_g(\vk)\delta_g(\vk')\>' \ =\ (b_\delta+\Delta b(k))^2P_{11}(k)+\vev{\epsilon_0^2}\, ,
\label{eq:Pgg1}
\end{align}
where $\vev{\epsilon_0^2}$ is the white noise arising from stochastic contributions in the bias expansion (see Sec.~\ref{sec:stoch}). 
On large scales, $k\to0$, the non-Gaussian contribution to the bias scales as $\Delta b(k) \propto k^{\Delta-2}$, which, for local non-Gaussianity ($\Delta=0$), recovers the classic result of Dalal et al.~\cite{Dalal:2007cu}. We see that the galaxy auto- and cross-correlation functions are boosted with respect to the dark matter correlation function for all $\Delta < 2$.

Note that equilateral PNG~($\Delta=2$) is not observable in this way, since on large scales $\Delta b(k)$ then approaches a constant which is degenerate with the Gaussian bias parameter~$b_\delta$.  One might think that the transfer function in (\ref{eq:deltab}) 
 introduces a scale dependence on smaller scales, $k \gtrsim k_{\rm eq} \approx 0.01\,\iMpch$, allowing $\Delta b$, in principle, to be distinguished from $b_\delta$.  However, for adiabatic perturbations, $M(k)$
can be expanded in powers of $k^2$, leading to a degeneracy with Gaussian
higher-derivative terms, such as $\nabla^2\d$.  
To estimate the size of the non-Gaussian contribution, let us assume that $b_\Psi$ depends on the small-scale initial fluctuations through the variance $\sigma_*^2$ on the scale $R_*$ (e.g.~for galaxies following a universal mass function, this would be the Lagrangian radius).  We then get~\cite{Schmidt:2012ys} 
\be
b_\Psi \simeq a_0 \frac{\partial\ln\bar n_g}{\partial\ln\sigma_*}\, \mu^2 R_*^2\,.
\ee
Moreover, we expect that $b_{\nabla^2\d}$ will involve the same nonlocality scale and thus be of order $R_*^2$.  
The scale dependence due to PNG with $\Delta=2$ is therefore larger than
that expected for the Gaussian higher-derivative terms, and thus detectable robustly, iff 
\be
|a_0 \fnl|  \,\gtrsim\, \left(\frac{\partial\ln\bar n_g}{\partial\ln\sigma_*}\right)^{-1} \,\frac{k_{\rm eq}^2}{\cH^2}
\,\simeq\, 10^3\:\left(\frac{\partial\ln\bar n_g}{\partial\ln\sigma_*}\right)^{-1} ,
\ee
where the final equality holds at redshift $z=0$.  Note that for
galaxies following a universal mass function, one has
$\partial\ln\bar n_g/\partial\ln\sigma_* = (b_\d-1)\d_c$, where $\d_c\approx 1.7$ is the spherical collapse threshold.  Hence, probing PNG with $\Delta=2$
robustly using the scale-dependent bias at levels below $\fnl$ of several hundred is 
not feasible due to degeneracies with Gaussian higher-derivative terms.  
This has so far not been taken into account in forecasted constraints for equilateral PNG (e.g.~\cite{Giannantonio:2012,Raccanelli:2015oma}).

\subsection{Bispectrum}
\label{sec:threepoint}

We now study the effects of the non-Gaussian terms in (\ref{eq:deltaNG}) on the galaxy bispectrum 
\be
B_g(k_1,k_2,k_3)\equiv \<\delta_g(\vk_1)\delta_g(\vk_2)\delta_g(\vk_3)\>'\, .
\ee
It will be convenient to write this as 
\begin{align}
B_g(k_1,k_2,k_3) \ =\ \  &b_\delta^3B_{111}(k_1,k_2 ,k_3)\nonumber\\[2pt]
&+\sum_{J\geq0}\left[P_{11}(k_1)P_{11}(k_2){\cal B}^{[J]}(k_1,k_2)\,\P_{J}(\hat\vk_1\cdot\hat\vk_2)+\text{2 perms}\,\right] .
\label{eq:Bg}
\end{align}
The first term corresponds to the linearly-evolved initial bispectrum, while the second term captures the nonlinear contributions (arising from both nonlinear gravitational evolution and nonlinear biasing), expressed in terms of dimensionless reduced bispectra ${\cal B}^{[J]}$.  We emphasize that (\ref{eq:Bg}) is valid in all momenta configurations, i.e.~it is \emph{not} restricted to the squeezed limit. To avoid confusion, we will use $J$ to denote the galaxy bispectrum expansion (\ref{eq:Bg}) and reserve $L$ for the Legendre expansion of the primordial squeezed bispectrum~(\ref{eq:FNLSLX}). Next, we will look at each multipole contribution $J$ in turn and determine the signatures of PNG in each of them.

\subsubsection*{Monopole} 

Let us first consider the non-stochastic contributions. 
At tree level, the operators $\delta$, $\Psi$, $\delta^2$ and $\Psi\delta$  contribute to the monopole ($J=0$) part of the galaxy bispectrum:
\begin{align}
{\cal B}^{[0]}(k_1,k_2) &= \big(b_\delta+\Delta b(k_1)\big)\big(b_\delta+\Delta b(k_2)\big)\bigg[\frac{34}{21}b_\delta+2b_{\delta^2} + \frac{b_{\Psi\delta}}{b_\Psi}\big(\Delta b(k_1)+\Delta b(k_2)\big)\bigg]\, ,
\label{eq:F0}
\end{align}
where $\Delta b(k)$ was defined in~(\ref{eq:deltab}). 
The first term in the square brackets is obtained by replacing $\delta_g(\vk_3)$ with the second-order solution of the linear term $\delta$, namely 
\begin{align}
\delta_{(2)}(\vk_3) &=
 \int\limits_{\vk_1}\int\limits_{\vk_2} \hat \delta_D(\vk_{1}+\vk_2-\vk_3)\bigg[\frac{17}{21}+\frac{1}{2}\P_1(\hat\vk_1\cdot\hat\vk_2)\left(\frac{k_1}{k_2}+\frac{k_2}{k_1}\right)\nonumber\\
 &\hskip 152.2pt+\frac{4}{21}\P_2(\hat\vk_1\cdot\hat\vk_2)\bigg]\delta_{(1)}(\vk_1)\delta_{(1)}(\vk_2)\, .
 \label{eq:delta2}
\end{align}
Of course, in (\ref{eq:F0}), we have only included the monopole part of the second-order solution, but we see that $\delta_{(2)}$ also contains a dipole and a quadrupole. 

\vskip 4pt
We also need to take into account the noise due to the stochastic nature of the relation between the galaxies and the underlying dark matter density. Given the results of Sec.~\ref{sec:stoch}, the noise contributions to the monopole of the bispectrum are 
\begin{align}
{\cal B}^{[0]}_{N}(k_1,k_2) \,=\ \, &\frac{\vev{\epsilon_0^3}}{3P_{11}(k_{1}) P_{11}(k_2)} \nonumber \\[2pt]
& + \left[(b_\delta+\Delta b(k_1))\left(\vev{\epsilon_0\epsilon_\delta}+\frac{\vev{\epsilon_0\epsilon_\Psi}}{b_\Psi}\Delta b(k_1)\right)\frac{1}{P_{11}(k_2)}+\{1\leftrightarrow2\}\right] .
\end{align}
 These terms are analytic in some of the external momenta. In position space, this will lead to terms proportional to Dirac delta functions. Note that these stochastic terms only affect the monopole of the bispectrum and, moreover, are absent when one considers the galaxy-matter-matter cross correlation.
\subsubsection*{Dipole} 
\label{ssec:dipole}
The leading contribution to the dipole ($J=1$) term in the galaxy bispectrum comes from 
the linear terms $\delta$ and $\Psi$ in the bias expansion. As discussed around~(\ref{eq:psiexp}), since the field $\Psi$ is  evaluated in Lagrangian space, it admits the following expansion around the Eulerian position $\vx$,
\be
\Psi(\vx,\tau) = \psi(\vx) + \boldsymbol{\nabla}\psi(\vx)\cdot\boldsymbol{\nabla}\Phi(\vx,\tau) + \cdots\, .
\ee
The second term in this expansion leaves an imprint in the dipole of the bispectrum.\footnote{Of course, by symmetry the dipole part of the galaxy bispectrum vanishes in the squeezed limit. However, as we emphasized before, the expansion (\ref{eq:Bg}) holds in all momentum configurations, so the dipole can be extracted away from the squeezed limit.}
Notice that the second-order solution $\delta_{(2)}$ in~(\ref{eq:delta2}) also contains a term proportional to~$\boldsymbol{\nabla}\delta_{(1)}\hskip -1pt\cdot\hskip -1pt\boldsymbol{\nabla}\Phi$ and therefore it also contributes to the dipole. Indeed, the total dipole contribution to the galaxy bispectrum~is
\be
{\cal B}^{[1]}(k_1,k_2)= \big(b_\delta+\Delta b(k_1)\big)\big(b_\delta+\Delta b(k_2)\big)\left[\frac{k_1}{k_2}\big(b_\delta+\Delta b(k_1)\big)+\frac{k_2}{k_1}\big(b_\delta+\Delta b(k_2)\big)\right] .
\label{eq:dipole}
\ee
We see that, even in the absence of PNG (i.e.~when $\Delta b(k)\to0$), the gravitational evolution leads to a dipole contribution. However, 
whenever the soft momentum scaling  of the bispectrum is less than two ($\Delta<2$),  the non-Gaussian contribution is enhanced on large scales relative to the Gaussian contribution. 
Interestingly, the contribution~(\ref{eq:dipole}) arises solely from the displacement induced by the velocity of the dark matter and the equivalence principle guarantees that no other operators aside from  $\delta$ or $\Psi$ yield this momentum dependence in the dipole. Hence, the dipole contribution is fully determined by the linear bias parameters which, in principle, can be measured in the galaxy two-point functions. It therefore serves as a consistency check for the scale-dependent bias $\Delta b(k)$ measured in the power spectrum. This is relevant since systematic effects can add spurious power to the large-scale galaxy power spectrum~(see e.g.~\cite{Pullen:2012rd}). The dipole of the galaxy bispectrum then is a useful diagnostic for determining whether the additional power measured in the power spectrum is due to a scale-dependent bias induced by PNG or arises from systematic effects which have not been accounted for.  On the other hand, for isotropic PNG with $\Delta=2$ (i.e.~PNG of the equilateral type), the dipole signature has the same degeneracy between $b_\d$ and the scale-independent $\Delta b$ as the auto and cross power spectra.  In fact, an effective scale-independent bias $b_\d^{\rm obs} = b_\d + \Delta b$ fitted to the large-scale power spectrum of galaxies will also be perfectly consistent with the measured large-scale bispectrum dipole in the case of $\Delta=2$.

\subsubsection*{Quadrupole}

Finally, we turn to the quadrupole contribution to the bispectrum. 
For isotropic ($J=0$) PNG this receives contributions from two terms: $(i)$~the second-order solution $\delta_{(2)}$ [cf.~(\ref{eq:delta2})] and $(ii)$~the square of the tidal tensor~$s_{ij}^2$. We find
\be
{\cal B}^{[2]}_{L=0}(k_1,k_2) =\frac{4}{3}\big(b_\delta+\Delta b(k_1)\big)\big(b_\delta+\Delta b(k_2)\big)\left[b_{s^2}+\frac{2}{7}b_\delta\right] .
\ee
As we have shown in \refssec{twopoint}, the linear bias $b_\delta$ and the scale-dependent bias $\Delta b(k)$ can be extracted from the galaxy power spectra by measuring the latter on a range of scales.  
If we can measure the quadrupole of the bispectrum over a similar range of scales, we can constrain the parameter $b_{s^2}$ and thus 
disentangle the contributions scaling as  $b_\d^2 \Delta b(k)$ and $b_{s^2} b_\d \Delta b(k)$.  Under the assumption of isotropic PNG, the quadrupole
can then be used to provide a second consistency test and further improve constraints on $\fnl$.

\vskip 4pt
In the presence of an anisotropic primordial squeezed limit  (with $L=2$), there is an additional contribution to the quadrupole of the galaxy bispectrum. In particular, the term $\Psi^{ij}s_{ij}$ in the bias expansion leaves the following imprint 
\begin{align}
{\cal B}^{[2]}_{L=2}(k_1,k_2) &=   \frac{b_{\Psi s}}{b_\Psi}\big(\Delta b(k_1)+\Delta b(k_2)\big)\big(b_\delta+\Delta b(k_1)\big)\big(b_\delta+\Delta b(k_2)\big)
\, .
\label{eq:anisoNG}
\end{align}
We see that if we wish to observe anisotropic non-Gaussianity through the scale-dependent bias, and disentangle it from the isotropic $L=0$ contribution, it is crucial that the bias parameters $b_\delta$ and $b_{s^2}$ are determined with enough precision to allow a measurement of the contribution~(\ref{eq:anisoNG}).  Thus, a measurement of the galaxy bispectrum over a range of scales is crucial.  It is also worth pointing out that if a scale-dependent bias is only observed through the quadrupole of the bispectrum, without a counterpart in the dipole or monopole (and also not in the power spectrum), 
this would prove the existence of a purely anisotropic primordial squeezed limit. 

\subsection{Stochasticity}

The results of this section have so far assumed the absence of any  large-scale stochasticity.  However, in general,
the galaxy statistics can receive contributions from stochastic terms (see~\refssec{stoch}).  In particular, when the primordial perturbations are produced by several fields during inflation, the short-scale fluctuations are also modulated by a field $\hat\Psi$ which is uncorrelated with the Gaussian long-wavelength fluctuations. We now discuss the signatures of such a term in the galaxy power spectrum and bispectrum. 

\subsubsection*{Power Spectrum}

In Sec.~\ref{sec:stoch}, we saw that a large collapsed limit of the four-point function introduces an additional stochastic term $\fnl b_{\hat\Psi} \hat\Psi$ in the bias expansion. This term is uncorrelated with long-wavelength fluctuations and only correlates with itself. Hence, it does not affect the galaxy-matter cross correlation but gives a non-vanishing contribution to the galaxy power spectrum
\be
P_{gg}(k)\,\supset\,  \fnl^2 b_{\hat\Psi}^2\,\<\hat\psi(\vk) \hat\psi(\vk') \>' \, .
\ee
Assuming $\Delta < 2$ for both $\psi$ and $\hat\psi$, the terms involving these fields will dominate on sufficiently large scales.  
In this case, the correlation coefficient between matter and galaxies in the large-scale limit becomes
\ba
r(k) \equiv \frac{P_{gm}(k)}{\sqrt{P_{gg}(k) P_{mm}(k)}}\,
\stackrel{\fnl\neq0}{=}
\, 
\frac{b_\Psi P_{1\psi}(k)}{\sqrt{[ b_\Psi^2  P_{\psi\psi}(k) +b_{\hat\Psi}^2 P_{\hat\psi\hat\psi}(k)] P_{11}(k)}}\,.
\label{eq:rk}
\ea
This is equal to unity if and only if $b_{\hat\Psi}=0$, otherwise the
correlation coefficient between matter and galaxies is less than one.  
Hence, by measuring the correlation coefficient between galaxies and
matter on large scales, we can determine whether the collapsed
limit of the four-point function exceeds the value predicted for initial conditions sourced by a single degree of freedom.  
Refs.~\cite{Tseliakhovich:2010kf,Baumann:2012bc} studied concrete models
in which this large-scale stochasticity arises.

\subsubsection*{Bispectrum}
Naturally, the stochastic term $\hat\Psi$ also affects the galaxy bispectrum. 
First, let us note that since the field $\hat\Psi$ is evaluated in Lagrangian space, it contributes a dipole to the bispectrum 
\be
{\cal B}^{[1]}_g(k_1,k_2)\, \supset\, b_\delta b^2_{\hat\Psi}\left(\frac{k_1}{k_2}\frac{P_{\hat\psi\hat\psi}(k_1)}{P_{11}(k_1)}+\frac{k_2}{k_1}\frac{P_{\hat\psi\hat\psi}(k_2)}{P_{11}(k_2)}\right) .
\ee 
As before, this dipole is fully determined by the parameters of the galaxy power spectrum, showing that the consistency relation between the power spectrum and the dipole part of the bispectrum is also valid in this case.  Of course, at the order at which we are working, one also needs to consider the effect of the operator $\hat\Psi\delta$. This will only contribute to the monopole part of the bispectrum
\be
{\cal B}^{[0]}_g(k_1,k_2) \, \supset\,b_\delta b_{\hat\Psi} b_{\hat\Psi\delta}\left(\frac{P_{\hat\psi\hat\psi}(k_1)}{P_{11}(k_1)}+\frac{P_{\hat\psi\hat\psi}(k_2)}{P_{11}(k_2)}\right) .
\ee
In particular, let us note that if $\hat\psi$ has the same scaling $\Delta$ as $\psi$, we have $P_{\hat\psi\hat\psi}(k)\propto [\Delta b(k)]^2P_{11}(k)$. Hence, this contribution can be comparable to (\ref{eq:F0}) in the regime where one or several momenta is small.  Finally, to be complete, we also need to account for the noise term $\epsilon_{\hat\Psi}\hat\Psi$ in the bias expansion. We find 
\be
{\cal B}^{[0]}_{N}(k_1,k_2) \, \supset\,b_{\hat\Psi}\<\epsilon_0\epsilon_{\hat\Psi}\>\frac{P_{\hat\psi\hat\psi}(k_1)+P_{\hat\psi\hat\psi}(k_2)}{P_{11}(k_1)P_{11}(k_2)}\, .
\ee
As before, the noise term only affects the monopole part of the bispectrum.

\section{Conclusions}
\label{sec:conclusions}

\begin{table}[t]
\begin{center}
\begin{tabular}{|l|c||c|c|c|c|c|}
\hline
 & & $P_g(k)$ & $r(k)$  & \multicolumn{3}{c|}{$B_g(k,k',k'')$} \\
type of PNG & $L$ & &   & monopole & dipole &  quadrupole \\
\hline
isotropic &  0 & $k^{\Delta-2}$ & -- & $k^{\Delta-2}$~* & $k^{\Delta-2}$ & $k^{\Delta-2}$~* \\
stochastic & 0 & $k^{\Delta-2}$ & $f(k,\Delta,\hat\Delta)$ & $k^{\Delta-2}$~* & $k^{\Delta-2}$ & $k^{\Delta-2}$~* \\
anisotropic & 2 & -- & -- & -- & -- & $k^{\Delta-2}$~* \\
\hline
\end{tabular}
\caption{Summary of the scale-dependent signatures of various types of PNG on the large-scale galaxy statistics: galaxy power spectrum, correlation coefficient with matter, and multipoles of the galaxy bispectrum.  Asterisks denote terms in the bispectrum that come with additional free parameters which need to be determined from smaller scales in order to constrain $\fnl$.  Note that in the stochastic case the scale-dependent bias and stochasticity in general have different scale dependences [see \refeq{rk}]. } \label{tab:summ}
\end{center}
\end{table}

In this paper, we have systematically investigated the impact of
primordial non-Gaussianity on the large-scale statistics of galaxies (or any other tracer of the large-scale structure).  We focused on the leading effects of quadratic non-Gaussianity on galaxy biasing, and provided a complete basis for the galaxy bias expansion to arbitrary order in perturbation theory.  
 The main effects depend on the momentum scaling of the squeezed limit of the primordial bispectrum, $(k_\ell/k_s)^\Delta$, and its angular dependence, $\P_L(\hat \vk_\ell \cdot \hat \vk_s)$.  Our findings are summarized in Table~\ref{tab:summ}.  
The different columns show the scale-dependent signatures in the galaxy
power spectrum $P_g(k)$, the correlation coefficient with matter $r(k)$, 
and the galaxy bispectrum $B_g(k,k',k'')$.  Our results for the two-point function
and the correlation coefficient recover previous results in the literature,
albeit arrived at in a more systematic way.   The bulk of the new results of this paper are contained in the galaxy bispectrum.  This bispectrum is naturally decomposed
into multipole moments; cf.~\refeq{Bg}. 
We showed that the dipole of the bispectrum allows
for a clean cross-check of the scale-dependent bias in the power spectrum,
without any additional free parameters.  The quadrupole of the bispectrum offers the possibility of constraining an anisotropic primordial bispectrum.  
The latter is generated in solid inflation~\cite{Endlich:2012pz} and in models with light additional spin-2 fields during inflation \cite{Arkani-Hamed:2015bza}.  

\vskip 4pt
Our systematic treatment allows for straightforward generalizations beyond
the leading PNG considered here:  
\begin{itemize}
\item \textit{Higher-order non-Gaussianity.}---The expansion in (\ref{eq:NGexp}) can
be continued to cubic and higher order, which corresponds to including the effects of a primordial
trispectrum and higher $N$-point functions.  We discuss these contributions in detail in Appendix~\ref{app:NG}.  
Generically these terms are small,
and can only be uniquely disentangled from lower-order non-Gaussianity when galaxy higher-point functions are measured on very large scales.  We therefore expect constraints from scale-dependent bias on higher-order PNG parameters to be significantly weaker than those on $\fnl$.  

\item \textit{Higher-spin non-Gaussianity.}---Note, however, that including higher-order non-Gaussianity and measuring higher $N$-point functions are essential in order to unambiguously constrain PNG with spin greater than two.  The two- and three-point functions are only sufficient for constraining spins 0 and 2.

\item \textit{Higher-derivative terms.}---Beyond the leading terms in the large-scale limit, we expect higher-derivative terms, such as $\nabla^2\Psi$, to appear in the bias expansion.  The scale determining the derivative expansion should be the same scale $R_*$ as for the Gaussian higher-derivative operators (e.g.~$\nabla^2 \d$).  
Note that for local-type PNG, the leading higher-derivative term will be
scale-independent, i.e.~it will appear as a very small correction 
to the Gaussian bias terms.  

\item \textit{Not-so-squeezed PNG.}---Beyond the squeezed limit, the primordial bispectrum receives $k_\ell/k_s$ corrections to its momentum scaling.  Through these corrections, biasing can in principle deliver additional information on the primordial bispectrum.  
However, disentangling PNG effects beyond the squeezed limit from the higher-derivative corrections to the bias expansion discussed above will be challenging.
\end{itemize}

Finally, it is important to emphasize that the considerations of this paper apply
specifically to the effects of PNG on \emph{biasing}.  Of course, galaxies
do retain the memory of non-Gaussianity in the initial conditions
by following the large-scale matter distribution; cf.~the term
$b_\delta^3 B_{111}$ in~(\ref{eq:Bg}).  Thus, in principle,
the galaxy three-point function does allow for a measurement of the
full bispectrum of the primordial potential perturbations beyond the
squeezed limit.  For this, it is crucial to include all relevant
operators in the bias expansion.  The results of this paper will thus be 
useful for
measurements and forecasts of constraints on PNG from large-scale structure.

\subsubsection*{Acknowledgements}

V.A.~and D.B.~thank
Daniel Green, Enrico Pajer, Yvette Welling, Drian van der Woude and Matias Zaldarriaga for collaboration on related topics.
D.B.~and V.A.~acknowledge support from a Starting Grant of the European Research Council (ERC STG grant 279617).  
V.A.~acknowledges support from the Infosys Membership.
F.S.~acknowledges support from the Marie Curie Career Integration Grant  (FP7-PEOPLE-2013-CIG) ``FundPhysicsAndLSS''.  

\appendix

\section{Systematics of the Bias Expansion}
\label{app:systematics}

In this appendix, we provide supplementary results on  \refsec{RHB}.  
Specifically, we present a Lagrangian basis of bias operators equivalent
to the Eulerian basis described in \refssec{basis}, and we extend the proof that the basis is closed under renormalization to all orders.  

\subsection{Lagrangian Basis}
\label{app:basis}

Consider a Lagrangian operator
$O_{\rm lgr}(\vq,\tau)$, for example a local operator made from $\partial_{q_i}\partial_{q_j}\Phi(\vq)$.  In perturbation theory, this operator can be written as
\be
O_{\rm lgr}(\vq,\tau) = D^{d_O}(\tau) O_{\rm lgr}^{[d_O]}(\vq)
+ D^{d_O+1}(\tau) O_{\rm lgr}^{[d_O+1]}(\vq) + \cdots\,,
\label{eq:Oexp}
\ee
where $D(\tau)$ is the growth factor and $d_O$ is the perturbative order of the leading contribution to $O_{\rm lgr}$.  The operators $O_{\rm lgr}^{[n]}$ are
constructed out of $n$ powers of second derivatives of the initial 
Lagrangian potential $\Phi(\vq)$, extrapolated to a given reference
epoch via linear growth.  In Lagrangian coordinates, convective time
derivatives reduce to simple time derivatives.  Allowing for time derivatives 
of $O_{\rm lgr}$ in the bias expansion is then
equivalent to including the contributions $O_{\rm lgr}^{[n]}$ at each
order individually.  This is because at any given order $N$, the
time derivatives are given by linear combinations of the terms $O_{\rm lgr}^{[n]}$ (this is most obvious when replacing $\tau$ with $\ln D(\tau)$ as the time coordinate).  Note that the statements regarding the nonlocality of
these operators made in \refssec{basis} also apply in Lagrangian space:  
even when starting with a local operator constructed out of 
$\partial_{q_i}\partial_{q_j}\Phi(\vq)$, the higher-order terms
$O_{\rm lgr}^{[n]}$ which are generated by time evolution are in general nonlocal, although
only comprising a small subset of all possible nonlocal operators.    
Only these specific nonlocal operators should be included in the operator
basis.  

To construct an explicit Lagrangian basis, we start with the Lagrangian distortion tensor,
\be
M_{ij} \equiv \frac{\der s_j}{\der q^i} \,, \quad \vx \equiv \vq + \v{s}(\vq,t)\, ,
\ee
and take all scalar contractions of $M^{[n]}_{ij}$ at 
each perturbative order.  The exception is ${\rm Tr}[M^{[n]}]$ with $n>1$,
which can be expressed in terms of lower-order operators through
the equations of motion \cite{Zheligovsky:2013eca}.  
For Gaussian initial conditions, the basis up to third order then is~\cite{Mirbabayi:2014zca}
\bea
{\rm 1^{st}} \ && \ {\rm Tr}[M^{[1]}]  \label{eq:listO} \\[3pt] 
{\rm 2^{nd}} \ && \ {\rm Tr}[(M^{[1]})^2]\,,\  ({\rm Tr}[M^{[1]}])^2 \nonumber\\[3pt] 
{\rm 3^{rd}} \ && \ {\rm Tr}[(M^{[1]})^3 ]\,,\ {\rm Tr}[(M^{[1]})^2]\hskip 1pt  {\rm Tr}[M^{[1]}],\ ({\rm Tr}[M^{[1]}])^3\,,\ {\rm Tr}[M^{[1]} M^{[2]}] \nonumber \, .
\eea
In the non-Gaussian case, we have to extend the basis by the field $\psi$, which is a nonlocal operator of the \emph{initial} density field.  At leading order in the non-Gaussianity, i.e.~to linear order in $\psi$, the non-Gaussian extension to the basis simply adds the field $\psi$ itself, as well as products of each operator of the Gaussian basis (\ref{eq:listO}) with $\psi$: 
\bea
{\rm  1^{st}} \ && \  \psi(\vq)  \label{eq:listONG} \\[3pt] 
{\rm 2^{nd}} \ && \ {\rm Tr}[M^{[1]}]\hskip 1pt \psi(\vq)  \nonumber \\[3pt]
{\rm 3^{rd}} \ && \ {\rm Tr}[(M^{[1]})^2] \hskip 1pt \psi(\vq)\,,\  ({\rm Tr}[M^{[1]}])^2 \hskip 1pt \psi(\vq) \, .\nonumber 
\eea
Finally, for anisotropic non-Gaussianity with $L=2$, we
have 
\bea
{\rm 1^{st}} \ && - 
\label{eq:listONGa} \\[3pt] 
{\rm 2^{nd}} \ && \ M^{[1]}_{ij} \psi^{ij}(\vq)  \nonumber \\[3pt]
{\rm 3^{rd}} \ && \ ({\rm Tr}[M^{[1]}]) \hskip 1pt M^{[1]}_{ij} \psi^{ij}(\vq) \, ,\nonumber
\eea
in exact analogy with (\ref{eq:listNGa}).

\subsection{Renormalization}
\label{app:renorm}

We now argue that the set of operators described in \refssec{basis}
and \refapp{basis} form a closed set under renormalization.  
In other words, all operators that appear in counterterms are already included in the general bias expansion,
so that these counterterms merely renormalize existing bias parameters.  This
is clearly a necessary condition for a self-consistent bias expansion.  

\vskip 4pt
In~\cite{Mirbabayi:2014zca} it was shown that the Gaussian basis
of operators constructed out of the operators $\Pi^{[n]}_{ij}$ is closed under renormalization.  Here, we will build on this result and show how
the additional non-Gaussian loop contributions can be related back to
Gaussian loop integrals.  

\vskip 4pt
Consider the renormalization condition (\ref{eq:RC}) for some operator $O$
that appears in the Gaussian bias expansion at some fixed order in perturbation theory:
\ba
C_{[O],m}^{\rm loop}(\vk,\vk_1,\cdots\hskip -1pt,\vk_m) \,\equiv\,  \vev{[O](\vk)\delta_{(1)}(\vk_1)\cdots\delta_{(1)}(\vk_m)}_{\rm loop}^{\prime} = 0\,.
\ea
A general loop contribution can then be written as
\ba
C_{O,m}^{\rm loop}(\vk,\vk_1,\cdots\hskip -1pt,\vk_m) \,\hat\d_D(\vk+\vk_{1\cdots m})
\,=\,\int_{\v{p}_1} \cdots\int_{\v{p}_n}& F_O^{(n)}(\v{p}_1,\cdots\hskip -1pt,\v{p}_n) \, \hat\d_D(\vk-\vp_{1\cdots n})
\label{eq:loopint}\\[3pt]
& \hspace{-1.5cm}\times 
\big\langle \d_{(1)}(\v{p}_1) \cdots \d_{(1)}(\v{p}_n)  \delta_{(1)}(\vk_1)\cdots\delta_{(1)}(\vk_m) \big\rangle\,, \nonumber
\ea
where $F_O^{(n)}$ is a generalized perturbation theory kernel specific to the operator $O$.  If the tree-level expression for $O$ starts at order $d$ 
(e.g.~$d=2$ for $O = \d^2$), then $F_O^{(n)}$ will have terms up to order $n-d +1$; i.e.~in terms of the standard SPT kernels $F_n$, it can involve various kernels up to $F_{n-d+1}$.  
By definition, $\d_{(1)}$ is the linearly-extrapolated \emph{initial} density field.  In the Gaussian case, we then use Wick's theorem to expand the $(n+m)$-point function into $(n+m)/2$ factors of the linear power spectrum $P_{11}$, if $n+m$ is even.  This leads to an $N$-loop contribution with
$N = n - (m+n)/2 = (n-m)/2$.  \refeq{loopint} vanishes if $n+m$ is odd.  

\vskip 4pt
Let us now consider the non-Gaussian case.  As in the main text, we will work to linear order in~$\fnl$,
and briefly comment on the generalization to higher-order PNG at the end. Moreover, for simplicity, we will restrict to spin-0 PNG.
For quadratic non-Gaussianity, as defined
in (\ref{eq:NGexp}), all higher $N$-point functions of the initial
conditions are obtained by inserting (\ref{eq:NGexp}) 
one or more times into Gaussian lower-point functions.  
At linear order in $\fnl$, we can therefore obtain any non-Gaussian loop
integral by one such insertion into a Gaussian integral (\ref{eq:loopint}),
formally increasing its loop order by one.  This yields terms of the following form
\begin{align}
& C_{O,m}^{\rm loop}(\vk,\vk_1,\cdots\vk_m) \,\hat\d_D(\vk+\vk_{1\cdots m})\: \supset \    
 \fnl 
\int_{\v{p}_1} \hskip -4pt \cdots\int_{\v{p}_n} 
F_O^{(n)}(\v{p}_1,\cdots \hskip -1pt,\v{p}_n) \,\hat\d_D(\vk-\vp_{1\cdots n})
 \label{eq:loopintNG}\\[4pt]
&
\quad \times \int_{\v{p}}  M(|\vp_i|)\, \Knl(\v{p},\v{p}_i-\v{p})\,
\big\langle \d_{(1)}(\v{p}_1) \cdots 
\varphi_\G(\v{p}) \varphi_\G(\v{p}_i-\v{p}) \cdots
\d_{(1)}(\v{p}_n)  \delta_{(1)}(\vk_1)\cdots\delta_{(1)}(\vk_m)
\big\rangle\, .
\nonumber \end{align}
Expanding the expectation value in the second line via Wick's theorem, leads to a non-zero contribution if $n+m$ is odd.  The dominant
contributions to the loop integrals are from the momentum regime $|\v{p}_j| \sim \Lambda$, where
$\Lambda$ is the cutoff of the integrals.  In this limit, we have
\be
\Knl(\v{p},\v{p}_i-\v{p})\varphi_\G(\v{p}) \sim 
\left(\frac{p}{\Lambda}\right)^{\Delta} \varphi_\G(\v{p})
\sim \left(\frac{p}{\Lambda}\right)^{\Delta} \left(\frac{\cH}{p}\right)^2 \d_{(1)}(\v{p})\, .
\ee
For $\Delta<2$, the relevant
non-Gaussian contribution therefore comes from the squeezed limit,
$|\vp| \ll |\v{p}_i|$, where the non-Gaussian kernel function is
\be
\Knl(\v{p},\v{p}_i-\v{p}) \xrightarrow{\ |\vp| \ll |\v{p}_i|\ }\left(\frac{p}{\mu}\right)^\Delta
\left(\frac{\mu}{p_i}\right)^\Delta\, . \label{eq:KNL}
\ee 
The expectation value in (\ref{eq:loopintNG}) then factorizes into
a two-point correlator involving $\varphi_\G(\v{p})$ and $\d_{(1)}(\vk_j)$,
where $j\in\{1,\ldots,m\}$, and
a correlator of $n+m-1$ factors of $\d_{(1)}$.  Setting
$j=m$ to be specific, we get
\begin{align}
& \left(\frac{p}{\mu}\right)^\Delta 
\vev{\varphi_\G(\v{p}) \delta_{(1)}(\vk_m)}' \,\times\, 
\left(\frac{\mu}{p_i}\right)^\Delta
\big\langle \d_{(1)}(\v{p}_1) \cdots 
\d_{(1)}(\v{p}_n)  \delta_{(1)}(\vk_1)\cdots\delta_{(1)}(\vk_{m-1})
\big \rangle'
\, . \label{eq:factNG}
\end{align}
We see that \refeq{loopintNG} is equivalent to one of the Gaussian 
loop integrals in \refeq{loopint} if we send $m \to m-1$ and make two further modifications.  First, one of the 
 power spectra in the loop is replaced by $P_{11}(p_i) \to (\mu/p_i)^\Delta P_{11}(p_i)$.  In practice, this means that the moments $\sigma^2(\Lambda)$ are
replaced by $\sigma_\Delta^2(\Lambda)$ as defined in (\ref{eq:sigmamod}).     
Second, the loop integral is multiplied by 
\be
\left(\frac{k_m}{\mu}\right)^\Delta M^{-1}(k_m) P_{11}(k_m)
= P_{ 1\psi}(k_m)\,,
\ee
where we have used the definition of $\psi$ in (\ref{eq:psidef})
 (the factor of $M^{-1}$ comes from relating $\varphi_\G$ to $\d_{(1)}$).  
In particular, for local-type non-Gaussianity ($\Delta=0$), any non-Gaussian 
$N$-loop integral is equivalent to a Gaussian $(N-1)$-loop contribution
with $m\to m-1$ and multiplied by $\fnl P_{1\varphi}(k_m)$.  
Hence, the simple remapping of loop corrections we found in \refssec{renorm}
continues to hold at higher loops as well.  This is not surprising, since we are only changing the
initial statistics by perturbing around the Gaussian case.  The gravitational
evolution is completely unchanged.  

Clearly, while conceptually simple, the non-Gaussian loop integrals of the 
form (\ref{eq:loopintNG}) are not absorbed by any terms in the Gaussian
bias expansion.  We therefore need to look at our extended non-Gaussian basis~(\ref{eq:listNG}).  
For every Gaussian operator $O$ there is now 
a non-Gaussian counterpart $\psi O$. 
We let $n$ and $m$ stand for the same values as in (\ref{eq:loopintNG}), such that $n+m$
is odd.  Consider then the specific loop contribution which 
involves the kernel $F_{\psi O}^{(n+1)}(\v{p}_1,\cdots\hskip -1pt,\v{p}_{n+1})$ for~$\psi O$.  
Since linear order in $\fnl$ is equivalent to linear order in $\psi$, 
this (unsymmetrized) kernel is simply proportional to $F_O^{(n)}(\vp_2,\cdots\hskip -1pt,\v{p}_{n+1})$.  
The loop integral for $\psi O$ then takes the following form 
\ba
C_{\psi O, m}^{\rm loop}(\vk,\vk_1,\cdots\hskip -1pt,\vk_m)
\,\hat\d_D(\vk+\vk_{1\cdots m})\: 
\supset\ &\int_{\v{p}_1} \cdots\int_{\v{p}_{n+1}} F_O^{(n)}(\v{p}_2,\cdots\hskip-1pt, \v{p}_{n+1}) \,\hat \d_D(\vk-\vp_{1\cdots n+1}) \label{eq:loopintNG2}\\[3pt]
& \ \times\big\langle \psi(\v{p}_{1}) \d_{(1)}(\v{p}_2) \cdots \d_{(1)}(\v{p}_{n+1})
\delta_{(1)}(\vk_1)\cdots\delta_{(1)}(\vk_m)
\big\rangle\,. \nonumber
\ea
Again, we assume that $\psi(\v{p}_{1})$ is only relevant if $|\v{p}_{1}|$ is 
set to one of the $k_j$ in the Wick expansion, corresponding to the
leading contribution on large scales, i.e.~we neglect terms
where $|\v{p}_{1}| \sim \Lambda$.  The expectation value then factorizes similarly to
\refeq{factNG}, with $\vp\to\v{p}_{1}$ and without the $(\mu/p_i)^\Delta$ prefactor.   
This shows that, apart from involving slightly different spectral moments, 
\refeq{loopintNG2} is of the same form as the loop integrals generated
by primordial non-Gaussianity.  We have thus shown that the basis of
operators (\ref{eq:listP}) and (\ref{eq:listNG}) is closed under renormalization
at linear order in $\fnl$.  

As we have emphasized repeatedly, for this to work it is essential that the field $\psi$
is defined in terms of the spatial position in the \emph{initial} conditions.  
Otherwise, the displacement from the initial to the final position in the argument
of $\psi$ would generate additional loop contributions which are not
canceled by any member of the basis $\left\{ \{O_{\rm G}\},\,\psi,\,\psi \{O_{\rm G}\}\right\}$, where $\{ O_{\rm G}\}$ stands for the basis for Gaussian initial conditions.   

The reasoning above extends to higher orders in $\fnl$,
provided that \refeq{NGexp} is the complete description of the non-Gaussian 
initial conditions at nonlinear order.  This then forces us to introduce
$\psi^2$ and $\psi^2 O_{\rm G}$ in the bias expansion.  Note, however, that
\refeq{NGexp} is not a generic expression beyond the three-point function
level.  We present a brief study of higher-order non-Gaussianity
in \refapp{higher}.  Furthermore, we point out that including subleading
terms in the squeezed limit expansion in the loop integrals will force
us to include additional higher-derivative terms as counterterms, as discussed at the end of \refssec{PNG}.  
  
Finally, exactly the same reasoning also
applies to higher-spin non-Gaussianity.  In that case, the auxiliary field
is a tensor, e.g.~$\psi_{ij}$ for $L=2$.  However, in the squeezed limit
this simply means that $\psi_{ij}$ has to be contracted with one of the external momenta
$\vk_m$, since terms with contracted loop momenta cancel by symmetry in the squeezed limit.    
Otherwise, the logic goes through as in the $L=0$ case.

\newpage
\section{General Primordial Non-Gaussianity}
\label{app:NG}

In this appendix, we outline how the results of the main text can be generalized to higher-order non-Gaussianity, and examine the conditions under which this leads to large-scale stochasticity.

\subsection{Higher-Order Non-Gaussianity}
\label{app:higher}

Scale-dependent bias arises due to a modulation of the
small-scale statistics of the initial density field $\d$ by a non-dynamical field $\psi$
which is related (in general, non-locally) to the initial potential~$\varphi$.  In this
section, we will deal exclusively with the initial conditions, so we 
drop the subscripts~$(1)$ to reduce clutter, i.e. $\d_{(1)}\to \d$.  The $N$-point functions of the density and the potential are related by
\be
\< \d(\vk_1,\tau) \cdots \d(\vk_N,\tau) \>' = M(k_1,\tau) \cdots M(k_N,\tau) \< \varphi(\vk_1) \cdots \varphi(\vk_N) \>'\,.
\ee
We can write
\be
\< \varphi(\vk_1) \cdots \varphi(\vk_N) \>' = A_N \left[ \Knl^{(N)}(\vk_1, \cdots,\vk_{N-1}) P_\varphi(k_1) \cdots P_\varphi(k_{N-1}) + {\rm perms}\, \right] ,
\ee
where $A_N$ denotes the dimensionless amplitude and $\Knl^{(N)}$ is the $N$-th order dimensionless kernel.  Throughout, all  $n$-point functions are connected.  
Note that as long as we include contributions from all $N$, there is no need to work beyond linear order in $A_N$ since those contributions can be absorbed by various $A_{N'}$ with $N' > N$.  We can then keep $N$ fixed and only consider one primordial $N$-point function.

\begin{figure}[t]
\centering
\includegraphics[width=0.45\textwidth, trim=1cm 19cm 0cm 2cm]{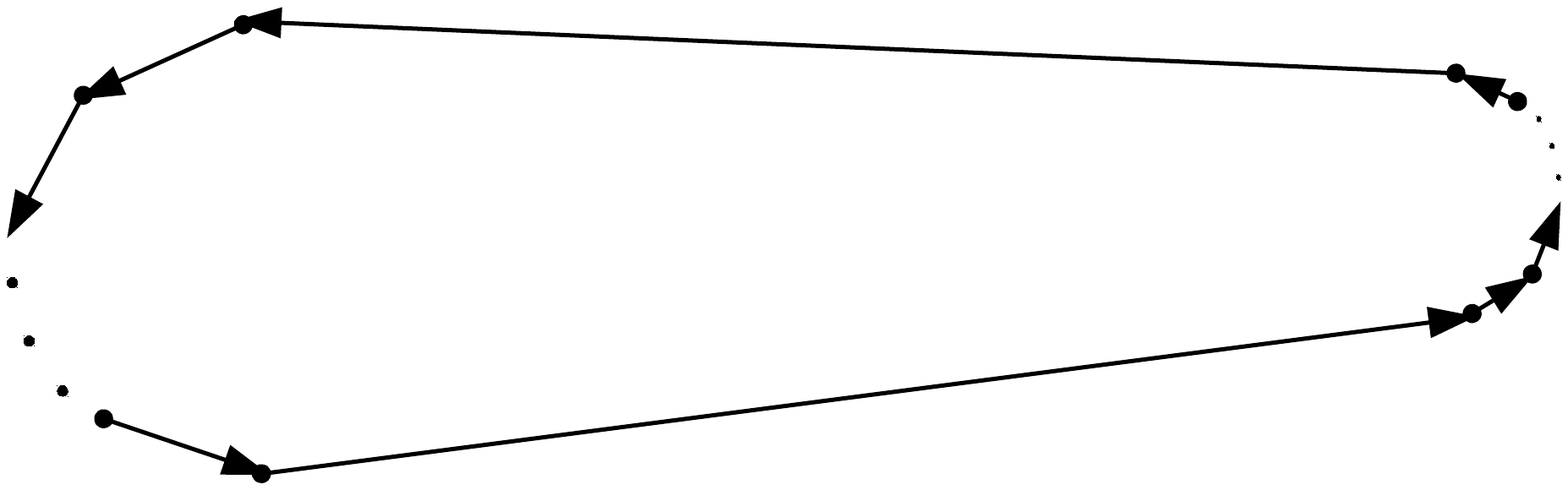}
\includegraphics[width=0.45\textwidth, trim=0cm 19cm 6cm 2cm]{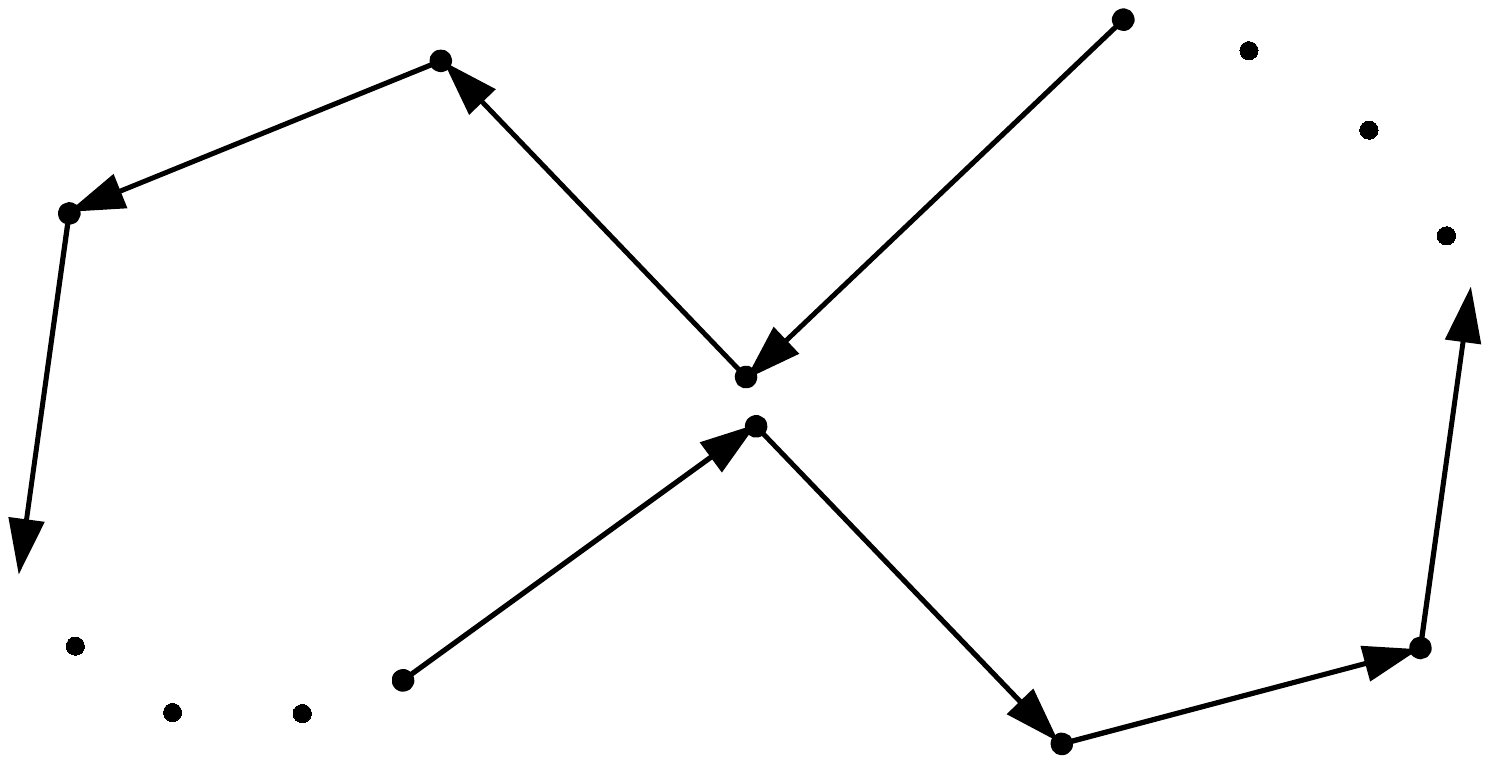}
\caption{Sketch of the ``generalized squeezed limit" for the case $n=2$ \emph{(left)}, and the ``collapsed limit" \emph{(right)}, considered in \refapp{higher} and \refapp{singlefield}, respectively.}
\label{fig:sketch}
\end{figure}

Consider $n$ small-scale modes $\vk_{s,1},\cdots\hskip -1pt, \vk_{s,n}$ ($2 \leq n < N$) and $N-n$ large-scale modes $\vk_1,\cdots \hskip -1pt,\vk_{N-n}$.  We will call this configuration the ``generalized squeezed limit'';
see \reffig{sketch} for an example with $n=2$.  This limit will affect  the large-scale $(N-n)$-point function of galaxies.  Note that this is different from the ``collapsed limit", where linear combinations of the short momenta $\vk_{s,i}$ combine to form large-scale modes, and which in general lead to stochasticity (see \refssec{stoch} and \refapp{singlefield}).  In the following, we will assume local-type PNG for simplicity, in which case the kernel functions $\Knl^{(N)}$ are simply constants, and we absorb these constants into rescaled amplitudes $A_N^{\rm loc}$.  The generalization to nonlocal PNG is straightforward, and mainly just involves additional factors of $(k_i/k_s)^\Delta$.  Furthermore, we will disregard factors of order unity, and in particular ignore the counting of permutations.  

We will only consider the leading contribution in the generalized squeezed limit, corresponding to the leading effect on LSS statistics on large scales.  This is given by  
\be
\< \d(\vk_1) \cdots \d(\vk_{N-n}) \d(\vk_{s,1}) \cdots \d(\vk_{s,n})\>' 
=  A^{\rm loc}_N F(k_1,\cdots\hskip -1pt,k_{N-n})  B^{\rm loc}_n(\vk_{s,1},\cdots\hskip -1pt, \vk_{s,n})\, , 
\label{eq:LSlimit}
\ee
where 
\begin{align}
F(k_1,\cdots\hskip -1pt,k_{N-n}) &\equiv \,M(k_1) P_\varphi(k_1) \cdots M(k_{N-n}) P_\varphi(k_{N-n}) \, ,\\
B^{\rm loc}_n(\vk_{s,1},\cdots\hskip -1pt, \vk_{s,n}) & \equiv M(k_{s,1}) \cdots M(k_{s,n}) \left[ P_\varphi(k_{s,1}) \cdots P_\varphi(k_{s,n-1})
+ {\rm perms} \right] .
\end{align}
Note that for $n=2$, we have $B^{\rm loc}_2 = P_{11}$.  For $n>2$, we
recognize  $B^{\rm loc}_n$ as the local-type primordial $n$-point function with $A^{\rm loc}_n=1$.   If we think of the small-scale modes as being measured in a patch over which the large-scale modes are effectively constant, then we can obtain \refeq{LSlimit} from the ansatz
\ba
\< \d(\vk_{s,1}) \cdots \d(\vk_{s,n})\>' \Big|_{\varphi(\vq)} = A^{\rm loc}_N \varphi^{N-n}(\vq) B^{\rm loc}_n(\vk_{s,1},\cdots\hskip-1pt, \vk_{s,n})\,,
\ea
which is a straightforward generalization of \refeq{Ploc}.  

We can now write down the terms that a primordial (local-type) $N$-point function will contribute to the bias expansion.  First, we need to parametrize the dependence of the local galaxy abundance $n_g(\xfl)$ on the local small-scale statistics at the Lagrangian position $\vq$.  One obvious choice are the (connected) moments of the density field smoothed on the Lagrangian scale $R_*$ of the galaxy,
\be
S_*^{(n)}(\vq) \equiv \< \d_*^n \>_{\vq} = \left(\hskip 1pt \prod_{i=1}^n \int_{\vk_{s,i}} \hskip -5pt W_*(k_{s,i}) \right)  
\< \d(\vk_{s,1}) \cdots \d(\vk_{s,n})\>_{\vq} \,,
\ee
where $W_*$ is the filter function.
We  find that $S_*^{(n)}$ is modulated by
\be
S_*^{(n)}(\vq) - \<S_*^{(n)}\> = A^{\rm loc}_N \varphi^{N-n}(\vq)  S_*^{(n),\rm loc}  
\, ,
\ee
where $S_*^{(n),\rm loc}$ is the $n$-th connected moment of the density field induced by local PNG of order~$n$ with $A^{\rm loc}_n = 1$, namely
\be
S_*^{(n),\rm loc} \equiv \left( \hskip 1pt \prod_{i=1}^n\int_{\vk_{s,i}} \hskip -5pt  W_*(k_{s,i}) \right)   B^{\rm loc}_n(\vk_{s,1},\cdots\hskip -1pt, \vk_{s,n}) \,
\hat\d_D(\vk_{s,1\cdots n})\, .
\label{eq:Slocdef}
\ee
  Defining $b_{\rm NG}^{(n)} \equiv \partial\ln \bar n_g/\partial S_*^{(n)}$ as the response of the galaxy abundance to a change in the $n$-th moment of the small-scale density field, we can then write down the contributions to the galaxy bias expansion 
\be
\d_g(\xfl) \supset A^{\rm loc}_N \sum_{m=1}^{N-2} \varphi^{m}(\vq) \hskip 1pt b_{\rm NG}^{(N-m)} S_*^{(N-m),\rm loc} \,.
\label{eq:biasNPNG}
\ee
Several points are worth noting about the result (\ref{eq:biasNPNG}):
\begin{itemize}
\item 
The operators to be
included in the bias expansion for higher-order local-type PNG are given by
powers of $\varphi(\vq)$.  Including gravitational evolution then yields the
basis $\left\{ O_{\rm NG} \right\} = \left\{ \{ O_{\rm G} \},\,\varphi^m,\,\varphi^m\{ O_{\rm G} \} \right\}_{m=1}^{N-2}$, where $\{ O_{\rm G}\}$
runs over the basis of operators for Gaussian initial conditions.  
\item Independently of the order of PNG, the relevant contributions to the galaxy power spectrum in the large-scale limit scale as 
\be
A^{\rm loc}_N P_{1\varphi}(k) \quad\mbox{and}\quad (A^{\rm loc}_N)^2 P_{\varphi\varphi}(k)\,.
\label{eq:contN}
\ee
This mean that no other scale dependences are generated apart from the well-known scalings $k^{-2}$ and $k^{-4}$.
\item Let us again consider the galaxy power spectrum.  For $N \geq 5$, the contributions with $m > 1$ merely renormalize the second term in (\ref{eq:contN}) and the first term if $m$ is odd; here, we have used that it is sufficient to work to linear order in $A^{\rm loc}_N$ in (\ref{eq:biasNPNG}), so that $\varphi(\vq)$ can be treated as Gaussian.  These terms only renormalize the non-Gaussian contributions to the galaxy power spectrum by a tiny amount 
proportional to powers of $\Delta_\varphi^2 = k^3 P_{\varphi\varphi}(k)/2\pi^2$ (unless, of course, the coefficients $A_N^{\rm loc}$ of higher-order PNG are enhanced by corresponding powers of $\Delta_\varphi^{-2}$).  
\item Given the previous point, it appears that the most relevant term in (\ref{eq:biasNPNG}), as far as the galaxy power spectrum is concerned, is the term with $m=1$:
\be
\d_g(\xfl) \supset A^{\rm loc}_N \varphi(\vq) \hskip 1pt b_{\rm NG}^{(N-1)} S_*^{(N-1),\rm loc} \, .
\ee
However, this contribution is also suppressed, as a simple estimate shows: Assuming that $b^{(N-1)}_{\rm NG}$ is of order one and taking all $k_s$ vectors to be of order $k_* \sim R_*^{-1}$ in (\ref{eq:Slocdef}), we find
\be
S_*^{(n),\rm loc} \sim \sigma_*^{2n-2} [M(k_*)]^{2-n} \sim \sigma_*^{2n-2} \left(\frac{k_{\rm eq}}{\cH}\right)^{4-2n}\,,
\ee
where $\sigma_*^2 \equiv \<\delta_*^2\>$ and in the last approximation we have assumed that $k_* \gg k_{\rm eq}$ at which point $M(k)$ saturates due to the transfer function.  For example, for $N=4$ (``$g_{\mathsmaller{\rm NL}}$''), the leading term corresponds to $n=3$ which, for  $\sigma_*$ of order unity, is suppressed by a factor of $(k_{\rm eq}/\cH)^{-2}  \sim 10^{-3}$.  In other words, in order to obtain the same amplitude of the scale-dependent bias, we need $g_{\mathsmaller{\rm NL}} \sim 10^3 \fnl$ (this has been verified in simulations \cite{Desjacques:2009jb,kendrick/etal:2012}).  
Hence, we find that, in the absence of a hierarchy between the amplitudes $A_N^{\rm loc}$, the scale-dependent bias due to higher-order PNG is highly suppressed.
\item We can generalize the above considerations to higher $n$-point functions of galaxies.  Bias operators involving $\varphi^m$ will only appear unsuppressed by either $\Delta_\varphi^2$ or
$S_*^{(n),\rm loc}$ in galaxy $n$-point functions with $n \geq m$.  Even then, they only
become relevant if at least $m$ of the $n$ wavenumbers $k$ are sufficiently small
(much smaller than $k_{\rm eq}$).  For example, the term $\propto\varphi^2$ is suppressed in the
squeezed limit of the galaxy bispectrum and only 
becomes relevant for three small $k$,
for which there is limited signal-to-noise.
\end{itemize}
We conclude that while the $n$-point functions of galaxies can in principle distinguish
PNG of various orders, the constraints on the amplitudes $A_N^{\rm loc}$ will weaken dramatically for $N>3$.  Furthermore, we note that the constraints on $g_{\mathsmaller{\rm NL}}\propto A_4^{\rm loc}$ obtained in \cite{Desjacques:2009jb} using the scale-dependent bias in the galaxy power spectrum are, unfortunately, completely degenerate with $\fnl$.  

\subsection{Collapsed Four-Point Function} 
\label{app:singlefield}

Up to cubic order, PNG generated by a single source can be written as
\begin{align}
\ph(\vk) = \ph_\G(\vk) &+ \int\limits_{\vp_1}\int\limits_{\vp_2} \hat \d_D(\vk-\vp_{12})\, \Knl(\vp_1,\vp_2)\ph_\G(\vp_1)\ph_\G(\vp_2)
\nonumber\\
&+\int\limits_{\vp_1}\int\limits_{\vp_2} \int\limits_{\vp_3}  \hat \d_D(\vk-\vp_{123})\, \Gnl(\vp_1,\vp_2,\vp_3)\, \ph_\G(\vp_1)\ph_\G(\vp_2)\ph_\G(\vp_3) - C\,, \label{eq:cubic}
\end{align}
where $C$ is a constant that ensures that $\vev{\varphi_\G}=0$.  This
constant will not play a role in the following and so we will drop it for clarity.  
We will assume that both kernels are non-singular for all kinematically allowed values of the momenta $\v{p}_i$. Specifically, in order for the $\Gnl$-contribution to the collapsed four-point function to be smaller than the $\Knl$-contribution, we assume that $\Gnl(\vk_1,\vk_2,\vk_3)\to (|\vk_{12}|/k_i)^\alpha$ as $|\vk_{12}/k_i|\to 0$ with  $\alpha>2\Delta-3$.\footnote{In principle, we could have $\fnl=0$ and still have a non-vanishing cubic interaction, which naively would seem to yield a stochastic component in the galaxy power spectrum. However, as discussed in App.~\ref{app:higher}, this effect is parametrized by a deterministic field (i.e.~a field which correlates with the long-wavelength degrees of freedom).}  
We now show that this ansatz leads to a definite prediction for the collapsed limit of the four-point function just in terms of the quadratic contribution.

\vskip 4pt
First, let us consider the contribution to the four-point function coming from the quadratic term in (\ref{eq:cubic}):
\ba
\hskip -5pt\<\ph(\vk_1)\ph(\vk_2)\ph(\vk_3)\ph(\vk_4)\> \,\supset&\, 
\int_{\vp_1}\int_{\vp_2}  \hat \d_D(\vk_1-\vp_{12})
\int_{\vp'_1}\int_{\vp'_2} \hat \d_D(\vk_2-\vp'_{12}) \nonumber\\[4pt] 
& \times \Knl(\vp_1,\vp_2) \Knl(\vp'_1,\vp'_2) \, \xi^{[6]}_{\varphi}(\vp_1,\vp_2,  \vp'_1,\vp'_2, \vk_3, \vk_4) + {\rm perms} \,,
\ea
where $ \xi^{[6]}_{\varphi}$ is the Fourier transform of the six-point function of $\varphi_\G$.
After applying Wick's theorem,
the momentum-conserving delta functions determine each of the momenta
$\vp_i,\,\vp'_i$, i.e.~there will be no loop integral left. In the following, we will consider the case of four comparable
small-scale modes~$\vk_i$.  One of the Wick
contractions is between one $\vp_i$ and one $\vp_j'$.  This leads to terms of the following form
\ba
\<\ph(\vk_1)\ph(\vk_2)\ph(\vk_3)\ph(\vk_4)\>' \,\supset\, 4 
\Knl(\vk_{13},-\vk_3) \Knl(\vk_{24}, -\vk_4) P_\ph(|\vk_{13}|) P_\ph(k_3) P_\ph(k_4)\,,
\ea
where the factor of 4 comes from the permutations $\vp_1 \leftrightarrow \vp_2$ and
$\vp'_1\leftrightarrow \vp'_2$, and we have written only one of several permutations of the momenta $\vk_i$.  In the collapsed limit,  e.g.~$|\vk_{13}|\ll k_i$, this can become a large contribution.  Let us denote $\vk_\ell \equiv \vk_{13}$ and hence $\vk_{24}=-\vk_\ell$.  We thus
have 
\begin{align}
\<\ph(\vk_1)\ph(\vk_2)\ph(\vk_3)\ph(\vk_4)\>' \ \ \stackrel{\rm collapsed}{\supset} \
\ &4\hskip 1pt
\Knl(\vk_\ell,-\vk_3) \Knl(-\vk_\ell, -\vk_4) \phantom{xxxxxxxxxxxxxxxxxx}\nonumber \\[2pt]
&\times P_\ph(k_\ell) P_\ph(k_3) P_\ph(k_4)\,.
\end{align}
This result is exactly what is expected from the modulation of the power spectrum in (\ref{eq:Ploc}):
using the fact that the collapsed limit of the four-point function  corresponds to correlating two small-scale power spectra, we have
\ba
\<\ph(\vk_1)\ph(\vk_2)\ph(\vk_3)\ph(\vk_4)\>' \ \ \stackrel{\rm collapsed}{\supset} \ \:&
\left \<P_\ph(\vk_3)\Big|_{\ph(\vk_\ell)} P_\ph(\vk_4)\Big|_{\ph(\vk_\ell')} \right\>  \nonumber \\[4pt]
=\:& 4\hskip 1pt  \Knl(\vk_\ell,\vk_3) P_\ph(k_3) \Knl(-\vk_\ell,\vk_4) P_\ph(k_4) P_\ph(k_\ell)\,.
\label{eq:4ptcoll}
\ea
To summarize, for the types of PNG that are described by a single degree of freedom, the quadratic contribution in~(\ref{eq:cubic}) leads to a definite relation between the squeezed limit of the bispectrum and the
collapsed limit of the four-point function: 
\ba
\<\ph(\vk_1)\ph(\vk_2)\ph(\vk_3)\ph(\vk_4)\>' \ \ \stackrel{\rm collapsed}{=}\ \:&
\frac{\<\ph(\vk_{13})\ph(\vk_3)\ph(\vk_3)\>' 
\<\ph(-\vk_{13})\ph(\vk_4)\ph(\vk_4)\>'}{4P_\ph(|\vk_{13}|)}   \,, \phantom{\,xx}
\label{eq:3pt4ptrel}
\ea
where
\be
\<\ph(\vk_{13})\ph(\vk_3)\ph(\vk_3)\>' =4\Knl(\vk_{13},\vk_3)P_\varphi(k_{13}) P_\varphi(k_3)\, . \label{eq:3ptsq} 
\ee
Next, we will show that this relation is unaffected by higher-order nonlinear terms.  

 The contribution to the four-point function coming from the cubic term in (\ref{eq:cubic}) is 
\ba
\<\ph(\vk_1)\ph(\vk_2)\ph(\vk_3)\ph(\vk_4)\> \ \supset \ &
\int\limits_{\vp_1}\int\limits_{\vp_2} \int\limits_{\vp_3}  \hat \d_D(\vk_1-\vp_{123})
 \, \Gnl(\vp_1, \vp_2, \vp_3) \,  \xi^{[6]}_{\ph}(\vp_1, \vp_2, \vp_3, \vk_2, \vk_3, \vk_4)  \vs
&  +\mbox{perms}\,. \label{eq:GNL4pt}
\ea
Again, we focus on the case of four comparable small-scale modes $\vk_i$.  One can easily see that there are no contributions from
large-scale modes to this four-point function.  Given our assumptions
stated after (\ref{eq:cubic}), the only possible
contribution arises from contracting say $\vp_1$ and $\vp_2$ in the six-point function, which yields
\be
\<\ph(\vk_1)\ph(\vk_2)\ph(\vk_3)\ph(\vk_4)\>' \propto 
\int_{\vp_1} P_\ph(p_1) \, \Gnl(\vp_1,-\vp_1,\vk_1)\, \xi^{[4]}_{\ph}(\vk_1,\vk_2,\vk_3,\vk_4)\,.
\ee
However, this term is absorbed by the renormalization of the
power spectrum that is necessary for a nonlinear relation of the
form (\ref{eq:cubic}).  Explicitly, we have
\be
\<\ph(\vk)\ph(\vk')\>' = P_\ph(k)\left[1 + 6 \int_{\vp} P_\ph(\vp) \,\Gnl(\vp,-\vp,\vk)\right] .
\ee
Absorbing this non-Gaussian correction into the renormalized $P_\ph(k)$ also absorbs the contribution of long-wavelength modes to the four-point function.   We therefore conclude that, as long as the non-Gaussian initial conditions are derived from a single stochastic variable,
the relation~(\ref{eq:3pt4ptrel}) is satisfied.

\vskip 4pt
Finally, let us emphasize that the collapsed four-point function (\ref{eq:4ptcoll}) and the squeezed 
bispectrum~(\ref{eq:3ptsq}) are exactly of the form necessary in order to
not generate any large-scale stochasticity as described in \refssec{stoch}.  
This is because both are derived from the same modulation of the power
spectrum (\ref{eq:Ploc}).  Following the reasoning in \refapp{higher},
at higher order, any particular collapsed limit of a primordial $N$-point
function will be similarly given by products of lower-order correlation functions.  
Hence, even for higher-order PNG, large-scale stochasticity
will not be generated as long as the primordial non-Gaussian field is sourced by a single
set of random phases.    

\newpage
\addcontentsline{toc}{section}{References}
\bibliographystyle{utphys}
\bibliography{refs}

\end{document}